\documentclass[12pt,epsfig]{article}
\usepackage{epsfig}
\usepackage{amssymb}

\begin{document}


\newcommand{\beq}{\begin{equation}}
\newcommand{\eeq}{\end{equation}}
\newcommand{\bea}{\begin{eqnarray}}
\newcommand{\eea}{\end{eqnarray}}
\newcommand{\beqn}{\begin{eqnarray}}
\newcommand{\eeqn}{\end{eqnarray}}
\newcommand{\beas}{\begin{eqnarray*}}
\newcommand{\eeas}{\end{eqnarray*}}
\newcommand{\defi}{\stackrel{\rm def}{=}}
\newcommand{\non}{\nonumber}
\newcommand{\bquo}{\begin{quote}}
\newcommand{\enqu}{\end{quote}}
\newcommand{\p}{\partial}


\def\de{\partial}
\def\Tr{ \hbox{\rm Tr}}
\def\const{\hbox {\rm const.}}
\def\o{\over}
\def\im{\hbox{\rm Im}}
\def\re{\hbox{\rm Re}}
\def\bra{\langle}\def\ket{\rangle}
\def\Arg{\hbox {\rm Arg}}
\def\Re{\hbox {\rm Re}}
\def\Im{\hbox {\rm Im}}
\def\diag{\hbox{\rm diag}}

\def\stroke{\vrule height8pt width0.4pt depth-0.1pt}
\def\topfleck{\vrule height8pt width0.5pt depth-5.9pt}
\def\botfleck{\vrule height2pt width0.5pt depth0.1pt}
\def\Zmath{\vcenter{\hbox{\numbers\rlap{\rlap{Z}\kern 0.8pt\topfleck}\kern
2.2pt\rlap Z\kern 6pt\botfleck\kern 1pt}}}
\def\Qmath{\vcenter{\hbox{\upright\rlap{\rlap{Q}\kern
3.8pt\stroke}\phantom{Q}}}}
\def\Nmath{\vcenter{\hbox{\upright\rlap{I}\kern 1.7pt N}}}
\def\Cmath{\vcenter{\hbox{\upright\rlap{\rlap{C}\kern
3.8pt\stroke}\phantom{C}}}}
\def\Rmath{\vcenter{\hbox{\upright\rlap{I}\kern 1.7pt R}}}
\def\Z{\ifmmode\Zmath\else$\Zmath$\fi}
\def\Q{\ifmmode\Qmath\else$\Qmath$\fi}
\def\N{\ifmmode\Nmath\else$\Nmath$\fi}
\def\C{\ifmmode\Cmath\else$\Cmath$\fi}
\def\R{\ifmmode\Rmath\else$\Rmath$\fi}


\def\QATOPD#1#2#3#4{{#3 \atopwithdelims#1#2 #4}}
\def\stackunder#1#2{\mathrel{\mathop{#2}\limits_{#1}}}
\def\stackreb#1#2{\mathrel{\mathop{#2}\limits_{#1}}}
\def\Tr{{\rm Tr}}
\def\res{{\rm res}}
\def\Bf#1{\mbox{\boldmath $#1$}}
\def\balpha{{\Bf\alpha}}
\def\bbeta{{\Bf\beta}}
\def\bgamma{{\Bf\gamma}}
\def\bnu{{\Bf\nu}}
\def\bmu{{\Bf\mu}}
\def\bphi{{\Bf\phi}}
\def\bPhi{{\Bf\Phi}}
\def\bomega{{\Bf\omega}}
\def\blambda{{\Bf\lambda}}
\def\brho{{\Bf\rho}}
\def\bsigma{{\bfit\sigma}}
\def\bxi{{\Bf\xi}}
\def\bbeta{{\Bf\eta}}
\def\d{\partial}
\def\der#1#2{\frac{\d{#1}}{\d{#2}}}
\def\Im{{\rm Im}}
\def\Re{{\rm Re}}
\def\rank{{\rm rank}}
\def\diag{{\rm diag}}
\def\2{{1\over 2}}
\def\ntwo{${\cal N}=2\;$}
\def\4N{${\cal N}=4$}
\def\none{${\cal N}=1\;$}
\def\x{\stackrel{\otimes}{,}}
\def\beq{\begin{equation}}
\def\eeq{\end{equation}}
\def\beqn{\begin{eqnarray}}
\def\eeqn{\end{eqnarray}}
\def\ba{\beq\new\begin{array}{c}}
\def\ea{\end{array}\eeq}
\def\be{\ba}
\def\ee{\ea}
\def\stackreb#1#2{\mathrel{\mathop{#2}\limits_{#1}}}

\def\baselinestretch{1.0}

\begin{titlepage}

\begin{flushright}
FTPI-MINN-05-48\\
UMN-TH-2420-05\\
\end{flushright}

\vspace{1cm}

\begin{center}

{\Large \bf Composite non-Abelian Flux Tubes
\\[2mm]  in \boldmath{$\mathcal{N}=2$} SQCD}
\end{center}

\vspace{0.5cm}

\begin{center}
{{\bf R.~Auzzi}$^{\,a}$, {\bf M.~Shifman}$^{\,a}$ and  {\bf A.~Yung}$^{\,a,b,c}$}
\end {center}
\begin{center}

$^a${\it  William I. Fine Theoretical Physics Institute,
University of Minnesota,
Minneapolis, MN 55455, USA}\\
$^b${\it Petersburg Nuclear Physics Institute, Gatchina, St. Petersburg
188300, Russia}\\
$^c${\it Institute of   Theoretical and Experimental Physics,
Moscow  117250, Russia}\\

\end{center}

\vspace{3mm}

\begin{abstract}
Composite non-Abelian vortices in $\mathcal{N}=2$ supersymmetric
$U(2)$  SQCD are investigated.
The internal moduli space of an elementary non-Abelian
vortex is $\mathbb{CP}^1$. In this paper we find a composite state
of two  coincident
non-Abelian vortices   explicitly solving the
 first order BPS equations. Topology of the internal moduli
space $\mathcal{T}$ is determined in terms of a discrete  quotient
$\mathbb{CP}^2/\mathbb{Z}_2$. 
The spectrum of physical strings and confined monopoles is discussed.
 This gives   indirect information about the sigma model with target
space $\mathcal{T}$.
\end{abstract}

\end{titlepage}

\section{Introduction}
\label{intro}

The Abrikosov vortex, also often referred to as the
Abrikosov--Nielsen--Olesen
(ANO) flux tube or the ANO string,
was one of the first important topological defects
discovered in field theory \cite{ANO}. It was also
one of the first Bogomolny completion  examples 
\cite{Bo}
which was later reinterpreted in supersymmetric setting as a BPS soliton 
\cite{Schaposnik}.
BPS saturation of the flux-tube-type solitons, such as the ANO string, is
due to the $(\frac{1}{2},\,\, \frac{1}{2})$ central charge \cite{GS} in the
underlying superalgebra.

The ANO string has two translational moduli characterizing the
position of the string center in the perpendicular plane.
In the supersymmetric case they are accompanied by two
supertranslational moduli. The effective low-energy theory on the
world sheet of the ANO string is trivial, it is a free field theory
of two bosonic moduli.

In the recent years it was realized that $\mathcal{N}=2$ $U(N)$ supersymmetric quantum chromodynamics
(SQCD) with the Fayet--Iliopoulos term supports a rich spectrum
of BPS solitons such as domain walls, $Z_N$ and non-Abelian strings,
monopoles, and their junctions, including boojums (for recent reviews see
\cite{rev1,rev2}).
In particular, the issue of BPS $Z_N$ strings was thoroughly discussed
and non-Abelian strings  discovered and analyzed 
 \cite{HT,vortici,tong-monopolo,SY-vortici,HT2,GSY,ssize,n=1*,
nonsusy,naw,ENS}.
(Abelian $Z_N$ strings were studied previously in \cite{VS}.)

In the theory with the $U(N)$ gauge group and $N_f=N$
flavors the solution for the elementary vortex 
displays, in turn, a rich structure: there are color-flavor 
locked zero modes  for the soliton solution, and the
resulting reduced moduli space is 
\[ \mathcal{M}=\mathbb{CP}^{N-1}.\]  
As discussed in Refs.~\cite{SY-vortici,HT2}
this property allows one to  directly connect
 the vortex solitons in the four-dimensional
$U(N)$ gauge theory with the $\mathbb{CP}^{N-1}$ sigma model in 
two dimensions. Moreover,
the kink of the $(1+1)$-dimensional theory is interpreted
as a BPS confined monopole located at the junction of two
magnetic strings \cite{tong-monopolo,SY-vortici,HT2}. 

In the $U(N)$ SQCD the ANO string is not minimal.
The tension of a ``minimal" string is $1/N$-th of that of the 
Abrikosov string. In particular, in the $U(2)$
model on which we will focus below the minimal string tension is
$2\pi\xi$ while the ANO string tension is $4\pi\xi$
where $\xi$ is a Fayet--Iliopoulos parameter (assumed to be positive).
Then it is natural to think of the ANO string as of composite, built of
two minimal strings. 
The question we will  address  in this paper is  the construction of
BPS composite flux tubes. We will limit ourselves to 2-strings,
introduce an appropriate ansatz and obtain, by a direct calculation,
a six-parametric family of solutions.

The Abrikosov string has only trivial translational moduli.
At the same time, 
if we consider two parallel minimal
(non-Abelian) strings  
at a distance $R$ from each other,
they are non-interacting because of their BPS nature,
and, if $R$ is large, we are certain that the configuration is characterized by
four internal moduli, in addition to two moduli
which have the meaning of the relative distance between
the  minimal strings. Thus, the reduced moduli space is six-dimensional.
How can one recover the Abrikosov string?

A constructive answer to this question will be given below.
The Abrikosov string will be shown to be
represented by  a singular point on the
moduli space of the 2-string.

In the general case  
the dimension of the $k$-string moduli space was
calculated \cite{HT} through the index theorem, $\nu = 2\,k\,N$.
This result has a clear-cut interpretation:
if the elementary vortices are taken at
large separations, the moduli space factorizes into
$k$ copies of $\mathbb{CP}^{N-1}$ plus the positions of
the elementary strings in the perpendicular plane;
each elementary string has two coordinates parameterizing its center.
Once the number of the collective coordinates is established at large separations
it stays the same at arbitrary separations. No potential can be generated on the moduli space because of   ``BPS-ness." In this respect the situation is  similar
to the BPS non-minimal ANO strings: the force due to the gauge 
boson exchange is canceled by the force due to the scalar Higgs fields, 
as can be checked by a direct calculation.

A general analysis of the geometry of the
six-dimensional  moduli space of the 2-string,
from a brane perspective, was carried out in  \cite{HT,HashT}.
It will be briefly reviewed in Sect.~\ref{twovortexbrane}.\footnote{
More precisely, in Ref.~\cite{HashT} the composite 2-string was
studied  through modeling the system in terms of string-theoretic
D-branes in the Hanany-Witten approach \cite{Hanany}.
The emphasis of \cite{HashT} was on scattering. See also Note Added.}

Our task is different: explicit construction of a family of the 2-string solutions
parametrized by a number of collective coordinates.
Unfortunately, we could not find a generic solution
with eight collective coordinates.
In this paper we present a six-parametric BPS solution for
the 2-string corresponding to the vanishing distance  $R$ 
between the elementary strings. Besides trvial translations, other
four collective coordinates 
present in our solution have the meaning of orientation
in the SU(2) group space. They will be referred to as internal moduli,
the corresponding moduli space being denoted by $\mathcal{T}$.
Thus, we construct a four-dimensional cross section of the
six-dimensional reduced moduli space (the reduced moduli space is obtained from
the full moduli space by factoring out overall translations.)

We find that the 
moduli space $\mathcal{T}$  is  
 given by a quotient
\beq
 \mathcal{T} \approx \mathbb{CP}^2/\mathbb{Z}_2\, . 
 \label{oneone}
\eeq
This result has a subtle distinction compared to the analysis of Ref.~\cite{HashT},
where the moduli space of two coincident strings was found to be
$\mathbb{CP}^2$. 
Our arguments supporting (\ref{oneone})
are collected in a systematic manner at the end of  Sect.~\ref{twovortexbrane}. 

While the metric of the 1-string sigma model is fixed
by symmetry arguments (it is the homogeneous metric in $\mathbb{CP}^1$ due to the
$SU(2)_{C+F}$ group, see below), the metric on the 2-string moduli space
is a much more complex object. In this issue we limit ourselves to a
general remark (Sect.~\ref{strms}) 
leaving this problem essentially open. 

On the other hand, the spectrum of confined monopoles
can be found in the Abelian limit $\Delta m \gg \Lambda$.
If we assume that the spectrum of   confined monopoles
does not change with $\xi$ as was the case for 1-strings
\cite{SY-vortici}, we
get an indirect information on the sigma model with the target space $\mathcal{T}$.

The paper is organized as follows. In Sect.~\ref{setup}
we briefly review our basic bulk theory, with the gauge group $U(2)$,
two flavors, and the Fayet--Iliopoulos term. 
Versions of this theory were consistently used as
a laboratory for various BPS solitons in the last few years.
In Sect.~\ref{three} we summarize aspects of the
Abelian  strings supported by the bulk theory
under consideration. Section \ref{four}
is devoted to non-Abelian elementary 1-strings.
 In Sect.~\ref{five} we thoroughly discuss the 2-string solution.
 Our basic ansatz is introduced in Sect.~\ref{ansatz}. We assemble BPS equations
 in Sect.~\ref{bpseq}. The numerical solution for the profile functions is
 presented in Sect.~\ref{numerical}, while the physical interpretation of the
 solution obtained is discussed in Sect.~\ref{physical}.
 We turn to the discussion of geometry of $\mathcal{T}$ in Sect.~\ref{strms}.
The issue of confined monopoles is addressed  in Sect.~\ref{seven}.
We summarize conclusions in brief in Sect.~\ref{eight}.
Appendices A and B deal with the zero modes of the
(1,1) and (2,0) strings, respectively.

\section{The Basic set-up and the Lagrangian}
\label{setup}

The bulk theory we work with has extended $\mathcal{N}=2$
supersymmetry,  $U(2)$ gauge theory
with $N_f=2$ matter hypermultiplets and a Fayet-Iliopoulos term $\xi$ for
 the $U(1)$ factor.
The following conventions are used:
\beqn
  \nabla_\mu &=&\partial_\mu-i \frac{\tau^a}{2} A_\mu^a -
\frac{i}{2} A^0_\mu \,,
\nonumber\\[3mm]
 A_\mu &=&
 \frac{\tau^a}{2} A_\mu^a + 
\frac{1}{2}A^0_\mu \,. 
\eeqn
The bosonic fields of the theory are the $U(2)$ gauge field, 
a zero charge scalar $a$, a complex adjoint scalar $a^a$ ($a=1,2,3$), the
fundamental scalars $Q^{kA}$ and $(\tilde{Q}^\dagger)^{kA}$ where 
$k=1,2$ is the color index of the SU(2) gauge subgroup and $A=1,2$
is the flavor index. We can 
write these last two fields as $2 \times 2$ matrices in the color-flavor 
indices $Q$ and $\tilde{Q}^\dagger$. The parameters of the theory
are the gauge couplings $e_0$ and $e_3$, the mass
parameters $m_A$ for each flavor and 
the Fayet-Iliopoulos term $\xi$. We can always consider the case in which the 
masses $m_A$ are real, while $\xi$ will be assumed to be positive.
Non-Abelian flux tubes emerge in the limit $m_1=m_2$.
It is convenient to start from $m_1\neq m_2$ (but keeping 
$|\Delta m| \equiv |m_1- m_2|\ll |m_{1,2}|$), in which case we will deal
with Abelian  strings, and then proceed to the
limit $m_1=m_2$.

The bosonic part of Lagrangian is
\beqn
{\cal L} &=& 
\int d^4  x \left\{\frac{1}{4 e_3^2} |F_{\mu \nu}^a|^2 +\frac{1}{4 e_0^2} 
|F_{\mu \nu}|^2+ 
 \frac{1}{ e_3^2} |D_{\mu} a^a|^2+ \frac{1}{ e_0^2}  |\partial_{\mu} a|^2 \right.
 \nonumber\\[3mm]
 &+& \left. \Tr (\nabla_\mu Q)^{\dagger} (\nabla_\mu Q)+ 
 \Tr (\nabla_\mu \tilde{Q}) (\nabla_\mu \tilde{Q}^{\dagger})+
 V(Q,\tilde{Q},a^a,a) \right\}
 \label{azione-tutta} 
 \eeqn
 where the potential $V$ is the sum of $D$ and $F$ terms,
 \beqn
V&=& \frac{e_3^2}{8}
 \left( \frac{2}{e_3^2} \epsilon^{abc} \bar{a}^{b} a^c +
 \Tr (Q^\dagger \tau^a Q) -\Tr(\tilde{Q} \tau^a \tilde{Q}^\dagger) \right)^2
 \nonumber\\[3mm]
  & +&   
 \frac{e_0^2}{8} \left(\Tr (Q^\dagger Q)-\Tr(\tilde{Q} \tilde{Q}^\dagger)-
2\xi\right)^2
 \nonumber\\[3mm]
  & +& \frac{e_3^2}{2} \left|\Tr (\tilde{Q} \tau^k Q) \right|^2 +
 \frac{e_0^2}{2} \left|\Tr (\tilde{Q}  Q) \right|^2
 \nonumber\\[3mm] 
  & +& \frac{1}{2} \sum_A \left|(a+\tau^b a^b+\sqrt{2} m_A) Q_A |^2+
 |(a+\tau^b a^b+\sqrt{2} m_A) \tilde{Q}^{\dagger}_{A} \right|^2 \,.
 \eeqn
Now, let us discuss the vacuum structure of our theory. The adjoint field 
vacuum expectation values
(VEVs) are
 \beqn
\langle a\rangle &=&-\sqrt{2}\, \frac{m_1+m_2}{2}\, ;  
\nonumber\\[3mm]
\langle a_3\rangle &=&-\sqrt{2}\, \frac{m_1-m_2}{2}=0\,,\quad\mbox{if}\quad
\Delta m = m_1-m_2 =0\,.
\label{avev}
\eeqn
If $m_1 \neq m_2 $, the gauge symmetry is broken to  $U(1)^2$ by the
VEV of the adjoint field.
Below we will consider mostly the 
case $\Delta m=0$ when the gauge group is not broken by 
the condensation of the adjoint field $a_3$.
The VEVs of the squark fields  are
\beq 
\langle \tilde{Q}\rangle =0; \qquad   \langle
Q\rangle
=\sqrt{ \xi }\left(\begin{array}{cc}
1 & 0\\
0 & 1 \\
\end{array}\right)\,.
\label{qvev} 
\eeq
The vacuum expectation value of $\langle Q\rangle$ completely   
breaks  the gauge symmetry, so that all gauge bosons
acquire masses in the bulk.

Note that if $\Delta m=0$, although both gauge and flavor groups are 
broken by the 
quark condensation, the {\em global} diagonal subgroup of the product of the 
gauge and flavor groups remains unbroken \cite{BaHa}. We call it SU(2)$_{C+F}$.
Its action on the quark fields is given by
\beq
Q\rightarrow U\,Q\,U^{-1},
\label{colorflavor}
\eeq
where the matrix $U$ on the left
corresponds to  the global color rotation while the matrix
$U^{-1}$ on the right is associated with  the flavor rotation. 
This mechanism is called  color-flavor locking.

With two matter hypermultiplets, the  SU(2) part of the gauge group
is asymptotically free,  implying generation of a dynamical scale 
$\Lambda$.
If descent to  $\Lambda$ were uninterrupted, the gauge coupling
$e_3^2$ would explode at this scale.
Moreover,  strong coupling effects in the SU(2) subsector at the
scale $\Lambda$ would break the  SU(2) subgroup through the
Seiberg-Witten mechanism \cite{SW1,SW2}.  Since we want to stay
at weak coupling   we assume
that 
\beq
\sqrt{\xi}\gg \Lambda\,.
\label{cond}
\eeq
This guarantees that the masses of all gauge bosons in the bulk
are much larger than $\Lambda$.

\section{Abelian strings}
\label{three}

Let us start from $\Delta m \neq 0$.
In this case the SU(2)$\times$U(1) group is broken to U(1)$\times$U(1)
by the VEV of the adjoint scalar field $a_3$, see Eq.~(\ref{avev}). 
Therefore, we have 
a lattice of Abelian strings labeled by two integers $(p,k)$
associated with winding with respect to two U(1) factors.
BPS strings  in the theory (\ref{azione-tutta})
were studied in \cite{MY}. Here we briefly review the main results 
of this paper.

\begin{figure}[h]
\epsfxsize=7cm
\centerline{\epsfbox{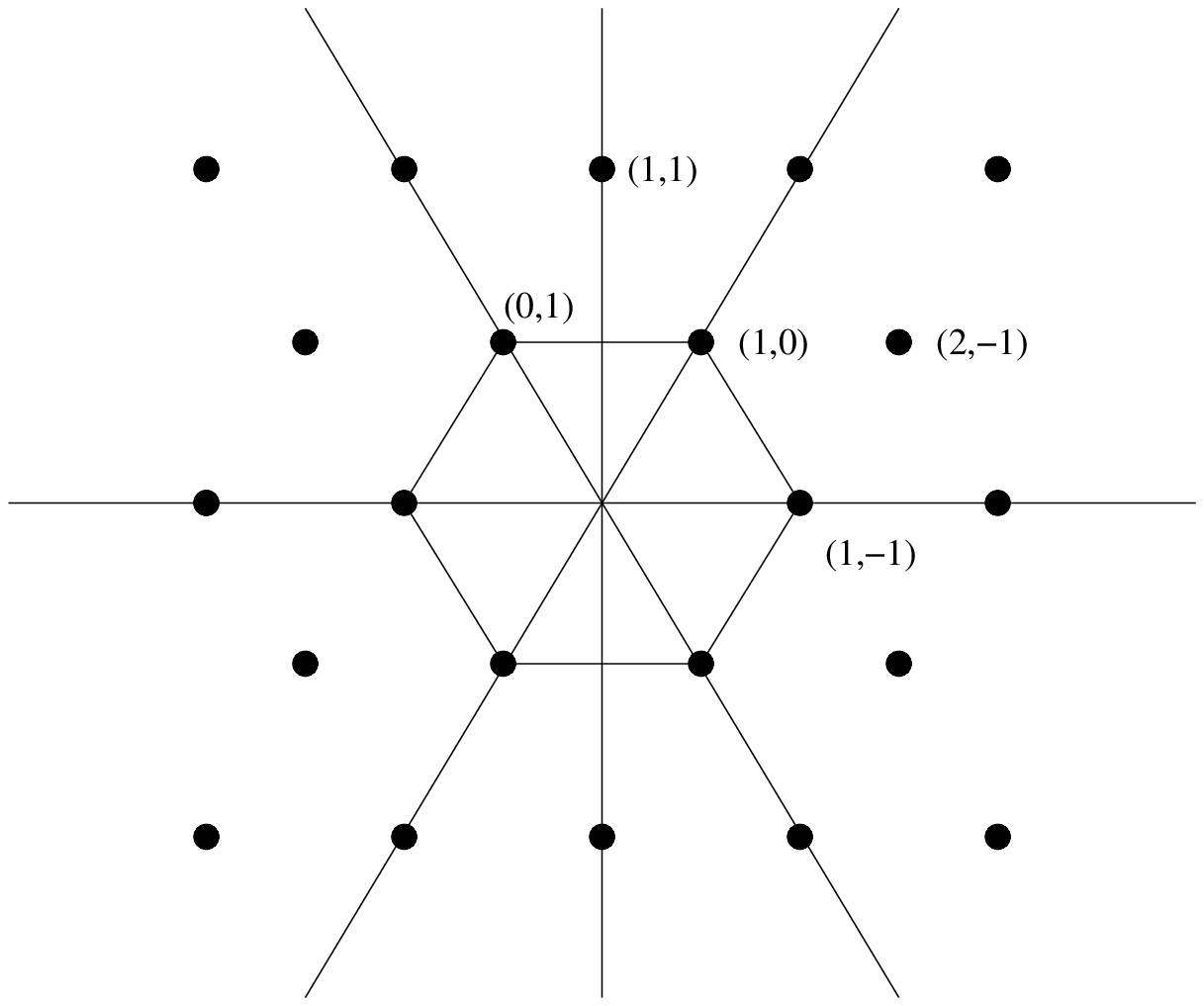}}
\caption{\footnotesize Lattice of $(p,k)$   vortices.    }
\label{fig:lattice}
\end{figure}

The charges of the $(p,k)$-strings can be plotted in the Cartan plane of
the  SU(3)  algebra. This is because our SU(2)$\times$U(1) gauge theory 
can be considered as a theory with the SU(3) gauge group broken down to
SU(2)$\times$U(1) at some high scale. Possible $(p,k)$-strings
form a root lattice of the SU(3) algebra \cite{MY}.
This lattice  is shown in Fig.~\ref{fig:lattice}. 
The vertical axis on this figure corresponds to charges with respect
to the U(1) gauge factor of the SU(2)$\times$U(1), while the horizontal axis
is associated with the $\tau^3$ generator of the SU(2) factor.

Two strings  $(1,0)$ and $(0,1)$ are   ``elementary'' or ``minimal"
BPS strings. They are often called $Z_2$ strings.
All other strings can be considered as  bound states of 
these elementary  strings, composites. If we plot two lines along the charges of these
elementary  strings (see Fig.~\ref{fig:lattice}) they divide the
lattice into four sectors. It turns out \cite{MY} that the strings
in the upper and lower sectors are BPS but they are marginally
unstable. On the contrary, the strings lying in the right and left sectors
are (meta)stable bound states of the  elementary ones;  they are 
{\em not} BPS-saturated.

The adjoint fields play no role in the string solutions. They are
equal to their VEVs (\ref{avev}). The same is true for the $\tilde{Q}$
quark: it vanishes on the string solution, which  is consistent with 
the equations of motion.  Hence, the  relevant part of the 
Lagrangian takes the following form:
\beqn
{\cal L}&\to&
\int d^4  x \left\{\frac{1}{4 e_3^2} \left|F_{\mu \nu}^a\right|^2 +\frac{1}{4 e_0^2} \left|F_{\mu \nu}\right|^2+ 
 \Tr (\nabla_\mu Q)^{\dagger} (\nabla_\mu Q)\right.
  \nonumber\\[3mm]
&+&\left. \frac{e_0^2}{8} (\Tr(Q^\dagger  Q)-2\xi))^2+
\frac{e_3^2}{8} (\Tr(Q^\dagger \tau^a Q))^2 \right\}.
\label{azione} 
\eeqn
This gives us an expression for the tension which, in the 
Bogomolny-completed
form \cite{Bo}, can be written as (the index $i,j=1,2$ run over 
the spatial coordinates
on the plane perpendicular to the string direction):
 \beqn
\label{bogomol}
T &=& \int{d}^2 x   \left (    \sum_{a=1}^3  \left[\frac1{2e_3}F^{(a)}_{ij } \pm
     \frac{e_3}{4}
\Tr \Big(Q^\dagger \tau^a Q\right)
\epsilon_{ij} \right]^2  
\nonumber\\[3mm]
 &+& \left[\frac1{2 e_0}F_{ij} \pm
     \frac{e_0 }{4 }
\left(\Tr (Q^\dagger Q)-2\xi \right)
\epsilon_{ij }\right]^2   
\nonumber\\[3mm]
& +&  \frac{1}{2} \left|\nabla_i \,Q^A \pm   i   \epsilon_{ij}
\nabla_j\, Q^A\right|^2
\pm
\xi  \tilde{F}
\Big)
\eeqn
Equating the non-negatively-defined terms in
the square brackets to zero gives us the first order
equations for the BPS strings. Then the last term in Eq.~(\ref{bogomol})
gives the string tension.
The   ansatz used to find   explicit
solution for the $(p,k)$-string  is 
\beqn 
Q&=& \sqrt{\xi} \left(\begin{array}{cc}
 e^{ip \varphi} \phi_1(r)& 0 \\
0 &  e^{ik \varphi}\phi_2(r) \\
\end{array}\right), 
\nonumber\\[4mm]
A^3_i
&=&
-\frac{\epsilon_{ij} x_j}{r^2} [(p-k)-f_3(r)]\,,\qquad 
A^0_i=-\frac{\epsilon_{ij} x_j}{r^2} [(p+k)-f(r)]\, ,
\label{astring}
\eeqn
where $\varphi$ and $r$ are polar coordinates in the 
perpendicular (1,2)-plane. The string axis is assumed to coincide with the $z$ axis.

Now, using the ansatz above,
the first order equations can be written  for the profile
functions $\phi_1,\phi_2,f,f_3$ \cite{MY,vortici,SY-vortici}, namely
\beqn
&& r\frac{\rm d}{{\rm d}r}\,\phi_1 (r)- \frac12\left(f(r)
+  f_3(r)\right)\phi_1 (r) =\ 0,
\nonumber\\[3mm]
&& r\frac{\rm d}{{\rm d}r}\,\phi_2 (r)- \frac12\left(f(r)
-  f_3(r)\right)\phi_2 (r) =\ 0,
\nonumber\\[3mm]
&& -\frac1r\,\frac{\rm d}{{\rm d}r} f(r)+\frac{e^2_0}{6}\,
\left(\phi_1(r)^2 +\phi_2(r)^2-2\xi\right)\ =\ 0,
\nonumber\\[3mm]
&& -\frac1r\,\frac{\rm d}{{\rm d}r} f_3(r)+\frac{e^2_3}{2}\,
\left(\phi_1(r)^2 -\phi_2(r)^2\right)\ =\ 0.
 \label{baba}
\eeqn
Furthermore, one needs to specify the boundary conditions
which would determine the profile functions in these equations. 
It is not difficult to see that  the appropriate boundary conditions are
\beqn
&&
f_3(0) = p-k\, ,\qquad f(0)=p+k\, ;
\nonumber\\[4mm]
&&
f_3(\infty)=0\, , \qquad   f(\infty) = 0
\label{fbc}
\eeqn
for the gauge fields, while the boundary conditions for  the
squark fields are
\beqn
\phi_1 (\infty)=\sqrt{\xi}\,,\qquad   \phi_2 (\infty)=\sqrt{\xi}\,,
\qquad \phi_1 (0)=0\, , \qquad \phi_2 (0)=0\, .
\label{phibc}
\eeqn
Numerical solutions to the
first-order equations (\ref{baba}) for the (0,1) and (1,0) elementary strings were
found in Ref.~\cite{vortici}. 
Numerical solutions for (2,0), (1,1) and (0,2) 2-strings will be presented in 
Sect.~\ref{numerical}.

The tension of the $(p,k)$-string is given by the boundary term in 
(\ref{bogomol}). We get
\beq
\label{ten}
T_{p,k}=2\pi\xi\,(p+k).
\eeq

\section{Non-Abelian 1-string}
\label{four}

\begin{figure}[h]
\begin{center}
\epsfxsize=3.5in \epsffile{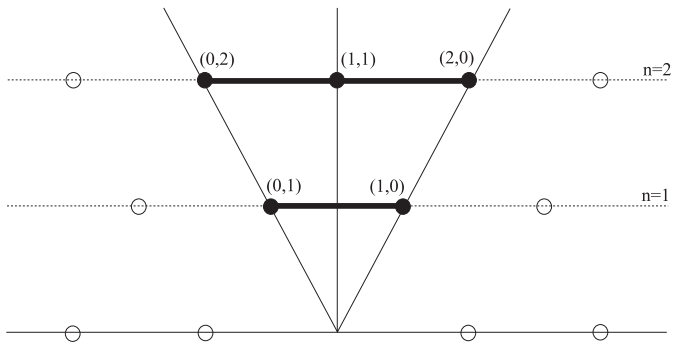}
\end{center}
\caption{\footnotesize Lattice of possible Abelian vortices.
In the non-Abelian case $m_1=m_2=m$ there is a moduli space interpolating
between different element of the lattice.}
\label{pallotino}
\end{figure}

In this section we review 
the elementary non-Abelian 1-vortex solution
which is associated with the elementary $(1,0)$ and $(0,1)$ Abelian strings
and emerges in the  limit $\Delta m=0$  \cite{vortici,SY-vortici}. If  $\Delta m =0$
the VEV of the adjoint scalar field $a_3$ does not break
the gauge group SU(2).
The relevant  homotopy group in this case is the fundamental group
\beq
\label{pi1suu}
\pi_1\left( \frac{{\rm SU}(2)\times {\rm  U}(1)}{ Z_2} \right) =  Z\, .
\eeq
This means that the $(p,k)$-string lattice reduces to a tower
labeled by a single integer  
\[ n=p+k,
\]
see Fig.~\ref{pallotino}. Note that the tension of all $(p,k)$-strings
with given $n$ are equal, see Eq.~(\ref{ten}).

For instance, the
$(1,-1)$ string becomes
classically unstable (no barrier).  On the SU(2) group manifold
it corresponds to a winding along the equator on the sphere $S_3$.
 Clearly this winding can be shrunk to zero
by contracting the loop toward the north or south poles 
of the sphere \cite{SYmeta}.
On the other hand, the elementary $(1,0)$ and $(0,1)$ strings
cannot be shrunk.
They correspond to a half-circle winding along the equator. The
$(1,0)$ and $(0,1)$ strings form a doublet of the residual global SU(2)$_{C+F}$.

A remarkable feature of the $(1,0)$ and $(0,1)$ strings is the
occurrence of non-Abelian moduli which are absent for the Abelian 
ANO strings. Indeed, while the vacuum field (\ref{qvev})
is invariant under the global SU(2)$_{C+F}$ (see Eq.~(\ref{colorflavor})),
the string configuration (\ref{astring}) is not.
Therefore, if there is a solution of the form
(\ref{astring}), there is in fact
a two-parametric family of solutions obtained from (\ref{astring}) by
the combined global gauge-flavor rotation.

In particular, for $(1,0)$ this gives
\beqn
&& Q^{kA}=\sqrt{\xi} U\left(
\begin{array}{cc}
 e^{  i \,  \varphi}   \phi_1(r) & 0  \\
  0 &  \phi_2(r) \\
  \end{array}\right)U^{-1} 
\nonumber\\[4mm]
&& =\sqrt{\xi}    e^{ \frac{i}{2} \, \varphi \,  (1+n^a\tau^a)} \,  U\left(
\begin{array}{cc}
  \phi_1(r) & 0  \\
  0 &  \phi_2(r) \\
  \end{array}\right)U^{-1}, 
  \nonumber\\[4mm]
&&  {\bf A}_{i}(x) = U  [- {\tau^3\o  2} \, \epsilon_{ij}\,\frac{x_j}{r^2}\,
[1-f_3(r)] ]   U^{-1} = -\frac12\,n^a \tau^a\epsilon_{ij}\,\frac{x_j}{r^2}\,
[1-f_3(r)], 
\nonumber\\[3mm]
&& A^{0}_{i}(x) = -\ \epsilon_{ij}\,\frac{x_j}{r^2}\,
[1-f(r)],
\label{1string}
\eeqn
where unit vector $n^a$ is defined by
\beq 
U \tau^3 U^\dagger=n^a \tau^a, \,\,\,\, a=1,2,3.
\label{na}
\eeq
Now it is   particularly clear that  this solution smoothly
interpolates between  the $(1,0)$ and
$(0,1)$ strings:   if $n=(0,0,1)$ the
first-flavor  squark  winds at infinity while for $n=(0,0,-1)$ it is the
second-flavor squark.

Since  the SU(2)$_{C+F}$  symmetry is not broken by the squark vacuum
expectation values,  it is physical and has nothing to do with
the gauge rotations eaten  by the Higgs mechanism. The orientational moduli
$ n^a$ are {\em not} gauge artifacts.
To see this  it is instructive to construct {\em gauge invariant} operators which
have explicit  $n^a$-dependence. Such a construction
is convenient in order to elucidate  features of our
non-Abelian string solution as well as for other purposes.

As an example, let us  define the
``non-Abelian" field strength,
\beq
 \tilde{\mathcal{F}}^a =\frac{1}{\xi}\,{\rm Tr}
\left(Q^\dagger F_3^{*b}\frac{\tau^b}{2}Q\, \tau^a
\right)\,,
\label{gidefi}
\eeq
where $F_k^{*}=1/2\varepsilon_{kij}F_{i,j}$ ($i,j,k=1,2,3$) and
the subscript 3 marks the $z$ axis, the direction of the string.
From the very definition it is clear that this field
is {\em gauge invariant}.\footnote{In the vacuum,
where the matrix $Q$ is that of VEV's,
$ \tilde{\mathcal{F}}^a$ and $ F_3^{*a}$ would coincide.}
Moreover,
Eq.  (\ref{1string}) implies that
\beq
 \tilde{\mathcal{F}}^a =- n^a\, \frac{(\phi_1^2+\phi_2^2)}{2\xi}\,
 \frac1r \,
\frac{df_3}{dr}\,.
\label{ginvF}
\eeq

From this formula we readily infer the physical meaning of the moduli
$ n^a$:
the flux of the {\em color}-magnetic field\,\footnote{Defined in the
gauge-invariant way, see Eq.~(\ref{gidefi}).} in the flux tube
is directed along $ n^a$.
For strings in  Eq.~(\ref{astring})  the
color-magnetic flux is
directed along the third axis in the SU(2) group space, either upward
or downward.
It is just this aspect that allows us to refer to the strings above
as ``non-Abelian."  

The internal moduli space of the vortex\,\footnote{In this case it coincides 
with the reduced moduli space obtained from
the full moduli space by removing overall translations.}
is given
by the symmetry group upon performing quotient with
respect to  the unbroken
part (in this case, the $U(1)$ subgroup generated by $\tau^a n^a$),
\beq
\tilde \mathcal{M}=SU(2)/U(1)=\mathbb{CP}^1=S^2.
\eeq
The vector $n^a$ is the coordinate in the moduli space $\tilde \mathcal{M}$.

An effective low-energy $(1+1)$-dimensional theory for the
vortex zero modes can be readily written (\cite{vortici,SY-vortici,HT}).
It turns out to be   an $\mathcal{N}=2$ $\mathbb{CP}^1$ sigma model
with the standard homogeneous metric. This is because
all non-translational  zero modes for the system are generated by the symmetry
$SU(2)_{C+F}$.

We will see that this is not the case for  2-strings
which, indeed, have additional zero modes not directly 
associated with the symmetry of the Lagrangian.
As it often happens, BPS solutions with higher topological charges
have more symmetry than the underlying Lagrangian.

\section{Non-Abelian 2-string}
\label{five}

\subsection{Preliminary remarks}
\label{prelrem}

If  $m_1 \neq m_2$ we return to the Abelian string situation.
The only solutions to Eqs.~(\ref{astring}) at level two (i.e. with $n=p+k=2$)
are the $(2,0)$,  $(1,1)$ and   $(0,2)$ strings.
In the  non-Abelian case ($\Delta m = 0$) we have the whole 
moduli space of solutions, with $(2,0)$, the $(1,1)$ and the $(0,2)$
strings being represented by particular points on this moduli space.

Let us first consider   two parallel elementary strings at a large separations,
$R=R_1-R_2\to\infty$. As soon as two strings  do not interact in this limit 
we conclude that the dimension of the moduli
space of this configuration is eight, twice the dimension of
the moduli space of each individual vortex. Two 
collective coordinates in this moduli
space correspond to the overall translations in the (1,2)-plane, 
two other collective coordinates
correspond to relative separations $R$, while the other four coordinates
are associated with the internal moduli space. At large $R$ the internal 
moduli space is 
 $\mathbb{CP}^1\times \mathbb{CP}^1$
( up to a discrete quotient, see Sect.~\ref{strms}) , described
by two orientational vectors $n^a_1$ and $n^a_2$ of the two constituent
strings.
Note that as soon as strings are BPS objects their interaction
potential vanishes, and the effective (1+1)-dimensional theory
on the string world sheet is a (classically) massless sigma model.

In this paper we obtain the 2-string solution at zero separation, $R=0$,
when both constituent strings are located at the same point in the 
(1,2)-plane,
i.e. are co-axial.
By continuity we expect that the internal moduli space is still
four-dimensional. 

Obtaining the four-parametric family
of solutions is  a serious problem. Suppose we 
start from the  (2,0) string solution, see (\ref{astring}) with $p=2$, $k=0$,
and apply rotation (\ref{colorflavor}) to this solution. Then we generate
only two-dimensional $\mathbb{CP}^1$ moduli space of solutions. In 
particular, this transformation interpolates only between the (2,0) and (0,2)
strings.

Moreover, the  (1,1) string imposes even a more severe problem.
The non-Abelian gauge potential is zero for this solution, and the matrix
$Q$ is diagonal, see Eq.~(\ref{astring}) at $p=k=1$. Therefore, the 
rotation (\ref{colorflavor}) acts on this solution trivially 
generating no internal moduli space at all. This can be viewed as a naive
embedding of the Abrikosov string.

Below we find the solution for the non-Abelian 2-string at $R=0$ by
explicitly
solving the first-order BPS equations. We show that the internal moduli
space is four-dimensional, as was expected. The family of solutions is described by
four parameters, one of them, $\alpha$, being the angle between two orientational vectors
$n^a_1$ and $n^a_2$ of two constituent strings.
At $\alpha=0$ and $\alpha= \pi$ the internal moduli space
develops singular throats, effectively reducing its dimension.
At $\alpha=0$ it becomes $\mathbb{CP}^1$ (the (2,0)/(0,2) string)
while for $\alpha=\pi$ (the (1,1) string) it shrinks to a point.

Our solution interpolates between all three Abelian strings: (2,0),
(0,2) and (1,1). To describe this solution
we introduce new profile functions which will depend on the polar coordinate
$r$ and, as a parameter, on the relative angle $\alpha$. 
The general BPS equations for the 2-string are then formulated in terms of these
profile functions.
Finding them at arbitrary $\alpha$ is a rather complicated calculation.
We perform an explicit analysis only
near particular points corresponding to the $(2,0)$ and  
$(1,1)$ vortices (presented in Appendices A and B).

\subsection{The ansatz}
\label{ansatz}

Our 2-string solution is parametrized by two vectors 
$\vec{n}_1$ and $\vec{n}_2)$.
The following expression is used for $Q$: 
\beq Q =\sqrt{\xi} \kappa(r) U_1 \left(\begin{array}{cc}
 z_1(r) e^{i \varphi}& 0 \\
0 & 1 \\
\end{array}\right)  U_1^{-1}
U_2  \left(\begin{array}{cc}
z_2(r) e^{i \varphi}& 0 \\
0 & 1 \\
\end{array}\right) U_2^{-1},
\eeq 
where
\beq U_1 \tau_3 U_1^{\dagger}=n_1^a \tau_a, \,\,\,\, U_2 \tau_3 U_2^{\dagger}=n_2^a \tau_a,\eeq
and $\kappa, z_1, z_2$ are functions of the radial coordinate $r$
and angle $\alpha$ between two vectors $n_1$ and $n_2$.
Taking $U_1=U_2=U_G$ the  global  orientational
zero modes are obtained. In order to study non-trivial $\alpha$-dependence
  we can  take
\beq 
\vec{n}_1 =(0,0,1),  \,\,\,\,
\vec{n}_2 = ( \sin \alpha, 0, \cos \alpha),
\label{vectorn}
\eeq
with $0\leq \alpha \leq \pi$.
Once the solution parametrized by the single parameter $\alpha$ is
found 
 we can recover
the general solution making a global rotation $U_G$.
In particular the functions $\kappa, z_1, z_2$
depend only on the relative angle $\alpha$ between 
$\vec{n}_1$ and $\vec{n}_2$ and not on the global orientation
of 2-string.

\begin{figure}[h]
\begin{center}
\epsfxsize=2in \epsffile{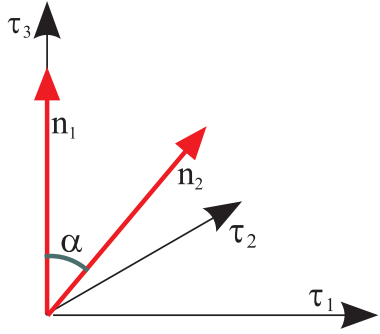}
\end{center}
\caption{\footnotesize It is always possible to align $\vec{n}_1$
with the $\tau_3$ axis and put $\vec{n}_2$ on the $\tau_3-\tau_1$ plane.
The angle between  $\vec{n}_1$ and $\vec{n}_2$ is $\alpha$. 
A global $SU(2)_{C+F}$ rotation introduced three extra angles.}
\label{axis}
\end{figure}

The particular choice (\ref{vectorn}) gives the 
following expression for $Q$ 
\beq 
Q = \sqrt{\xi} \kappa \left(\begin{array}{cc}
 z_1 e^{i \varphi}& 0 \\
0 & 1  \\
\end{array}\right) 
U  \left(\begin{array}{cc}
z_2 e^{i \varphi}& 0 \\
0 & 1 \\
\end{array}\right) U^{-1}
\ ,
 \label{zuma}
\eeq
where
\beq
U  \left(\begin{array}{cc}
z_2 e^{i \varphi}& 0 \\
0 & 1 \\
\end{array}\right) U^{-1} = \frac{(z_2 e^{i \varphi}+1)}{2} {\bf 1}
+  \frac{(z_2 e^{i \varphi}-1)}{2} (\vec{\tau}\cdot\vec{l})
\eeq 
and 
\beq 
\vec{l} = ( \sin \alpha,\,\,0 ,\,\,
\cos \alpha)\,.
\eeq
A more explicit expression for $Q$ has the form
{\small
\beq Q = \sqrt{\xi} \kappa \left(\begin{array}{cc}
\left(\cos^2  \frac{\alpha}{2} \right)   e^{2 i \varphi}  z_1 z_2+
 \left(\sin^2\frac{\alpha}{2}\right) e^{ i \varphi}  z_1 
& \frac{\sin \alpha}{2}  (e^{ 2 i \varphi}  z_1 z_2
 -e^{  i \varphi} z_1) \\[3mm]
\frac{\sin \alpha}{2}  
(e^{ i \varphi}  z_2 -   1 )
& \left(\cos^2 \frac{\alpha}{2} \right)  
+\left( \sin^2 \frac{\alpha}{2}\right) e^{ i \varphi} z_2 \\
\end{array}\right),
\eeq}
where $\varphi$ is the polar angle.
The BPS equations  are
\beq 
(\nabla_1+i\nabla_2)\, Q=0\,,
\eeq
which can be identically rewritten as
\beq 
A_1+iA_2= - i (\partial_1 Q +i \partial_2 Q)\,  Q^{-1}\,. 
\label{redbaron}
 \eeq
Substituting the ansatz (\ref{zuma})  in this expression
gives us the form of the gauge fields.
 The result of a rather tedious  calculation is
\beqn
 && - i (\partial_1 Q +i \partial_2 Q) Q^{-1}=  i e^{i \varphi} 
\left( \frac{2}{r}-2\frac{\kappa'}{\kappa}
 -\frac{z_1'}{z_1}-\frac{z_2'}{z_2} \right){\mathbf 1}
  \nonumber\\[3mm]
 && +i e^{i \varphi} \left( \frac{1+\cos \alpha}{r}-\frac{z_1'}{z_1}
 -\cos \alpha \frac{ z_2'}{z_2} \right) \tau_3   
 \nonumber\\[3mm]
 &&
+e^{i \varphi} (\sin \alpha) \left(
\frac{1}{r}-\frac{z_2'}{z_2}
 \right) \left(i \frac{z_1^2+1}{2  z_1} (\cos\varphi)-
  \frac{z_1^2-1}{2  z_1} (\sin \varphi)
 \right) \tau_1 
 \nonumber\\[3mm]
 && +e^{i \varphi} (\sin \alpha) \left(
\frac{1}{r}-\frac{z_2'}{z_2}
 \right) \left(-i \frac{z_1^2+1}{2  z_1} (\sin \varphi)-
  \frac{z_1^2-1}{2  z_1}( \cos \varphi)
 \right) \tau_2\,.  \nonumber\\
 \label{gaufie}
 \eeqn
 In order to satisfy Eq. (\ref{redbaron}) we choose the following gauge potentials:
\beqn
&& A^0_{(i)} = -\frac{\epsilon_{ij} x_j}{r^2}  (2-f)   \,,
\nonumber\\[3mm]
 &&  A^3_{(i)} = -\frac{\epsilon_{ij} x_j}{r^2}  ((1+\cos \alpha )-f_3) \,,
 \nonumber\\[3mm]
 && 
 A^1_{(i)} = -\frac{\epsilon_{ij} x_j}{r^2} (\sin \alpha) (\cos\varphi)  (1-g)
-  \frac{ x_i}{r^2}  (\sin \alpha) (\sin \varphi)  h \,,
\nonumber\\[3mm]
 && 
 A^2_{(i)} = +\frac{\epsilon_{ij} x_j}{r^2} (\sin \alpha) (\sin \varphi)  (1-g)
-  \frac{ x_i}{r^2}  (\sin \alpha)( \cos \varphi)  h \,.
\nonumber\\
\label{f1}
\eeqn
To facilitate reading, let us summarize here our set of the
profile functions. The set includes
\beq
\kappa\,,\quad z_i\,\,\, (i=1,2)\,,\quad f\,, \,\,\,f_3\,, \,\,\, g\,, \,\,\, h\,.
\eeq
Now we calculate the field strength tensor 
\beq  
F_{\mu\nu}=\partial_\mu A_\nu-\partial_\nu A_\mu -\frac{i}{4} 
[A_\mu^a \tau^a,A_\nu^b \tau^b ]. 
\label{fst}
\eeq
Note that the commutator term does not vanish now, while in 
 the 1-string case it was zero. Technically this is a very important distinction.

The only non-vanishing component of the field strength is
$F_{(12)}^a$, namely,
\beqn
F^0_{(12)} &=& -\frac{f'}{r} \,,
 \nonumber\\[3mm]
F^3_{(12)} &=& -\frac{f'_3}{r} +\frac{(1-g) h (\sin \alpha)^2}{r^2} \,,
 \nonumber\\[3mm]
F^1_{(12)} &=& (\cos \varphi) (\sin \alpha) \left( -\frac{g'}{r}
-\frac{\cos \alpha-f_3}{r^2} h  \right) \,,
 \nonumber\\[3mm]
F^2_{(12)} &=& - (\sin \varphi) (\sin \alpha) \left( -\frac{g'}{r}
-\frac{\cos \alpha-f_3}{r^2} h  \right). 
\label{f2} 
\eeqn

\subsection{The BPS equations}
\label{bpseq}

The full set of the BPS equations we will deal with are
\beqn
&& \tilde{F}^a_{(3)}+\frac{e_3^2}{2}\Tr (Q^{\dagger} \tau^a Q)=0\,,
  \nonumber\\[3mm]
&& \tilde{F}^0_{(3)}+\frac{e_0^2}{2}(\Tr (Q^{\dagger}  Q)-2 \xi)=0\,,
 \nonumber\\[3mm]
&&  A_1+iA_2= - i (\partial_1 Q +i \partial_2 Q) \, Q^{-1}\, . 
\label{fuse}
\eeqn
Substituting our ans\"atze we get the
following system of the first-order differential equations:
\beqn
&& 
\frac{f'}{r}=\frac{e_0^2}{4} 
\left((1+z_1^2)(1+z_2^2)\kappa^2+
 \cos \alpha(1-z_1^2)(1-z_2^2)\kappa^2-4\right)\,,
 \nonumber\\[4mm]
&& \frac{f'_3}{r}   - \frac{(1-g) h (\sin \alpha)^2}{r^2}  
 =\frac{e_3^2}{4} 
\left((z_1^2-1)(z_2^2+1)\kappa^2+
 \cos \alpha (z_1^2+1)(z_2^2-1)\kappa^2\right)\,,
\nonumber\\[4mm]
&& \frac{g'}{r}+\frac{\cos \alpha-f_3}{r^2} h =\frac{e_3^2}{2} \kappa^2 z_1 (z_2^2-1)\,,
\nonumber\\[4mm]
&& \frac{f}{r}= 2\frac{\kappa'}{\kappa}+\frac{z_1'}{z_1}+\frac{z_2'}{z_2}\,,
\nonumber\\[4mm]
&& \frac{f_3}{r}= \frac{z_1'}{z_1}+
\cos \alpha \frac{z_2'}{z_2} \,,
\nonumber\\[4mm]
&& \frac{1-g}{r}= \frac{z_1^2+1}{2 z_1}
 \left(\frac{1}{r}-\frac{z_2'}{z_2}\right)\,.
 \label{afsub}
\eeqn
The function $h$ can be expressed in terms of   other profile functions,
\beq 
h = \frac{z_1^2-1}{z_1^2+1} (1-g)\, .
 \label{afoth}
\eeq
The  boundary conditions that must be imposed on 
the profile functions at $r\rightarrow 0$ are
\beqn
&& f(r)=2+\mathcal{O}(r^2), \qquad  f_3(r)=(1+\cos \alpha) +
\mathcal{O}(r^2)\,, 
 \nonumber\\[2mm]
&&  g(r)=1+\mathcal{O}(r^3),  \qquad  h(r)=\mathcal{O}(r^3)\,,
 \nonumber\\[2mm]
&& z_1(r)\to  \mathcal{O}(r),\qquad z_2(r)\rightarrow \mathcal{O}(r), \qquad \kappa(r)\rightarrow 
\mathcal{O}(1)\,.
 \label{bczero}
\eeqn
The boundary conditions  at $r\rightarrow \infty$ are
\beqn
&& f,\,f_3\,,g\,,h \rightarrow  0\,, 
 \nonumber\\[2mm]
&& \kappa,\,z_1,\,z_2 \rightarrow 1\,.
 \label{bcoo}
\eeqn

We see that  the boundary conditions for the gauge profile functions
$f$ and $f_3$ at $r=0$ are 
\beq
f(0)=2\quad\mbox{ and }\quad f_3(0)= 1+\cos\alpha\, . 
\eeq
This is in accordance with the  boundary conditions for the
Abelian strings, Eq.~(\ref{fbc}).
For the (2,0) string we have $p=2$, $k=0$, and  Eq.~(\ref{fbc}) gives 
$f(0)=2$, $f_3(0)= 2$. This corresponds to $\alpha=0$ in 
Eq.~(\ref{bczero}); the
vectors $n_1^a$ and $n_2^a$ of two 1-string constituents of the 2-string
are parallel. 

For the (1,1) string we have $p=1$, $k=1$, and Eq.~(\ref{fbc})
gives $f(0)=2$, $f_3(0)= 0$. This case corresponds to $\alpha =\pi$ in 
Eq.~(\ref{bczero}), so that the
vectors $n_1^a$ and $n_2^a$ are anti parallel. 

\subsection{Another gauge}
\label{another}

With an appropriate gauge transformation (only a constant color rotation,
no flavor rotation)
\beq U=\exp\left(i \tau_2 \left(\pi-\frac{\alpha}{2}\right)\right),
\eeq
we can cast the solution in the following form:
 \beq 
 Q = \sqrt{\xi} \kappa \left(\begin{array}{cc}
-\cos \frac{\alpha}{2}  e^{ 2 i \varphi}  z_1 z_2
& \sin \frac{\alpha}{2}  e^{  i \varphi}  z_1  \\[3mm]
- \sin \frac{\alpha}{2}  e^{  i \varphi}  z_2 
& -\cos \frac{\alpha}{2}  \\
\end{array}\right).
\eeq 
Then the gauge field takes the form
\beqn 
&& A_\varphi=\left(\begin{array}{cc}
\frac{-3-\cos \alpha+f+f_3}{2r}  &  \frac{e^{i \varphi}(1-g) \sin \alpha}{2r} \\[3mm]
\frac{e^{-i \varphi}(1-g) \sin \alpha}{2r}
&\frac{-1+\cos \alpha+f-f_3}{2r} \\
\end{array}\right),
\nonumber\\[3mm]
&& A_r=\left(\begin{array}{cc}
0  &i e^{i \varphi}  \frac{\sin \alpha}{2r} h \\[3mm]
-i e^{-i \varphi}  \frac{\sin \alpha}{2r} h
& 0  \\
\end{array}\right).
\eeqn
In this gauge the expressions are more compact;
the VEV of the squark filed $Q$ at  infinity takes the form
\beq 
Q = \sqrt{\xi}  \left(\begin{array}{cc}
-\cos \frac{\alpha}{2}  e^{ 2 i \varphi} 
& \sin \frac{\alpha}{2} e^{  i \varphi}   \\[3mm]
- \sin \frac{\alpha}{2}  e^{  i \varphi}  
& -\cos \frac{\alpha}{2}  \\
\end{array}\right).
\eeq 

\subsection{Numerical Solution}
\label{numerical}

\begin{figure}[h]
\begin{center}
$\begin{array}{c@{\hspace{.2in}}c@{\hspace{.2in}}c} \epsfxsize=1.5in
\epsffile{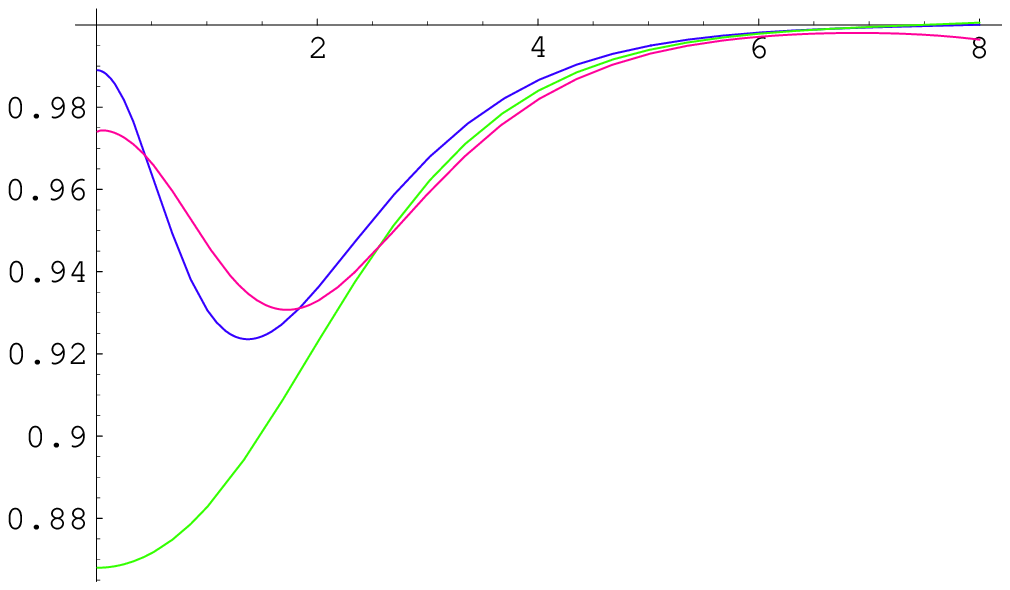} &
    \epsfxsize=1.5in
    \epsffile{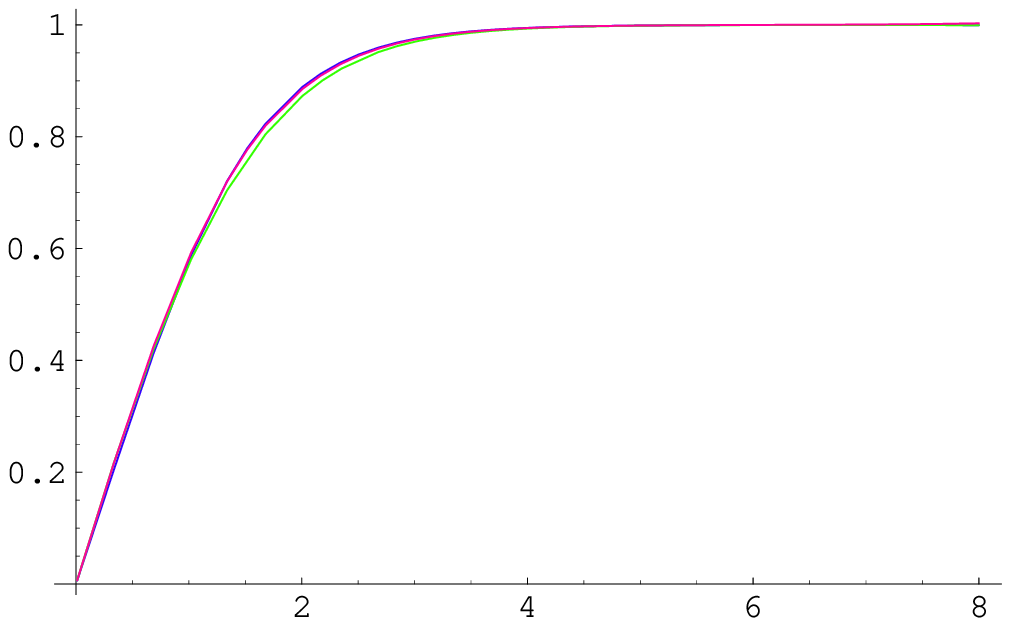} &
     \epsfxsize=1.5in
    \epsffile{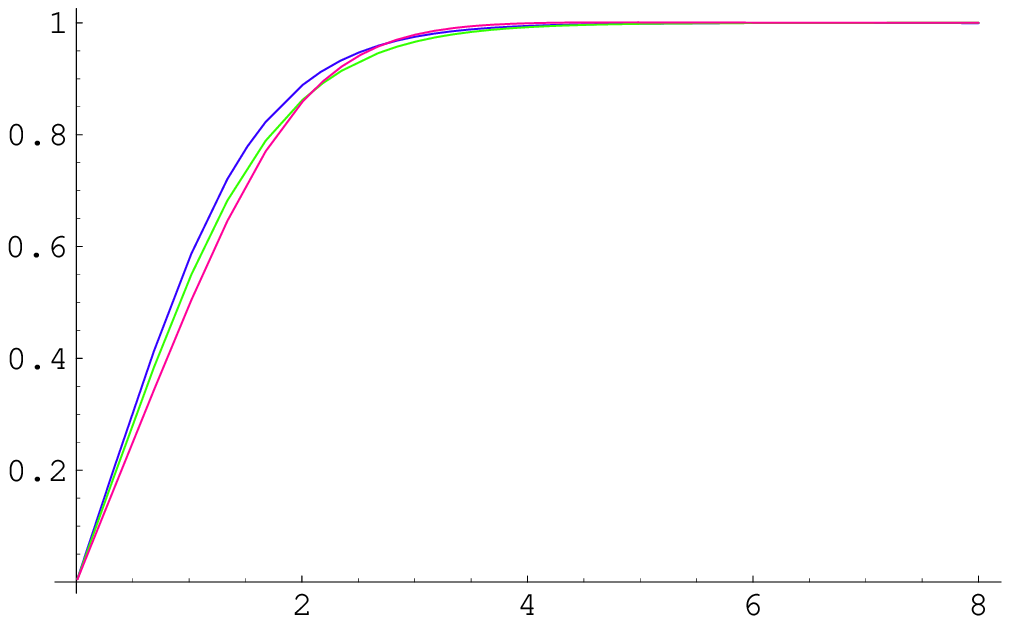}
\end{array}$
\end{center}
\caption{\footnotesize  $\kappa(r)$ (left):
$z_1(r)$ (center), $z_2(r)$ (right), 
at $\alpha=0$ (green), $\alpha=\frac{\pi}{2}$ (red), $\alpha=\pi$ (blue).
} \label{confronto1}
\end{figure}

\begin{figure}[h]
\begin{center}
$\begin{array}{c@{\hspace{.2in}}c@{\hspace{.2in}}c} \epsfxsize=1.5in
\epsffile{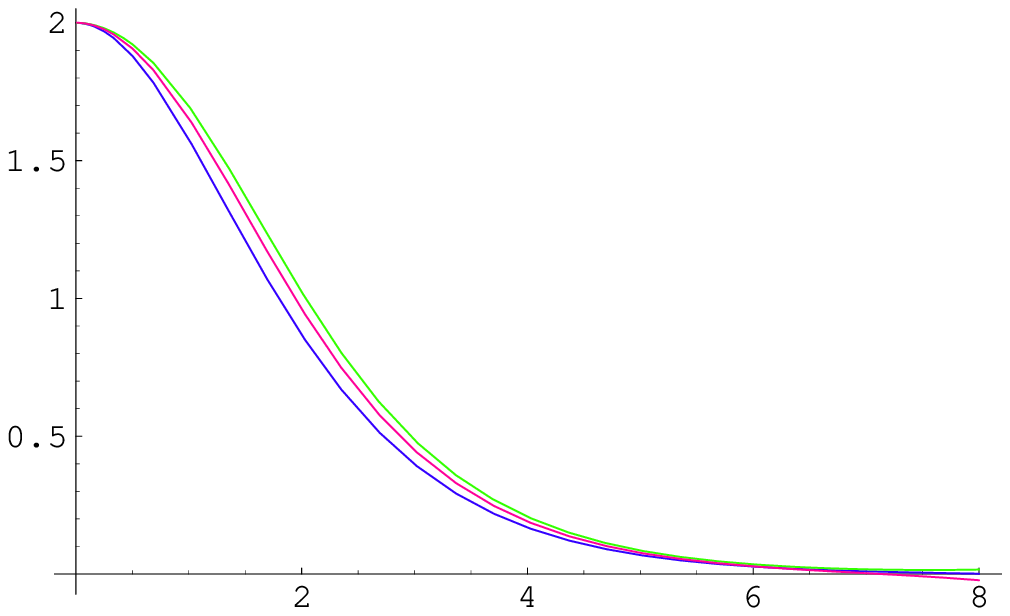} &
    \epsfxsize=1.5in
    \epsffile{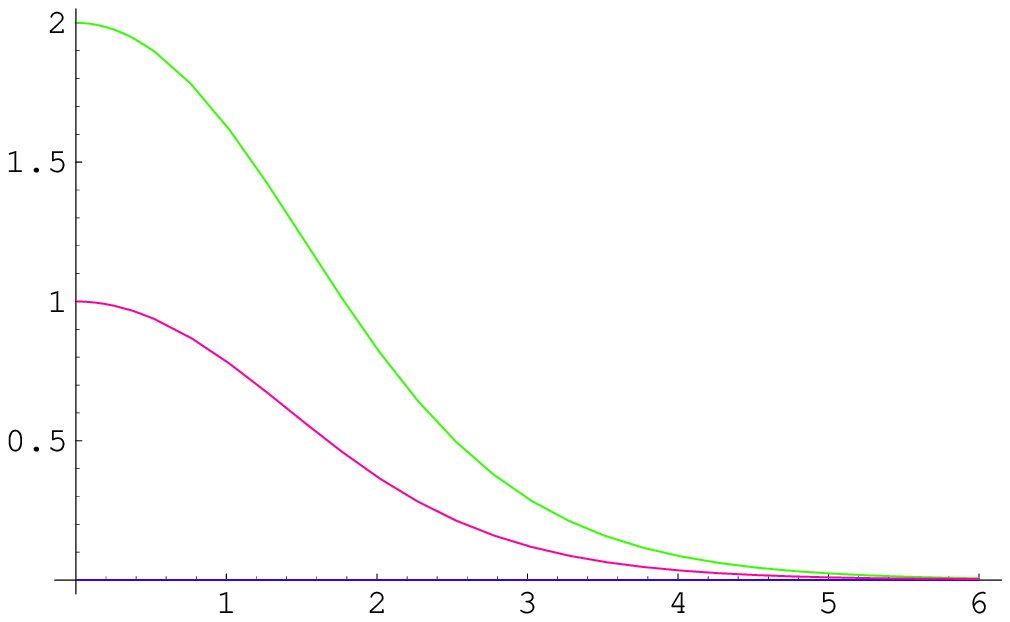} &
     \epsfxsize=1.5in
    \epsffile{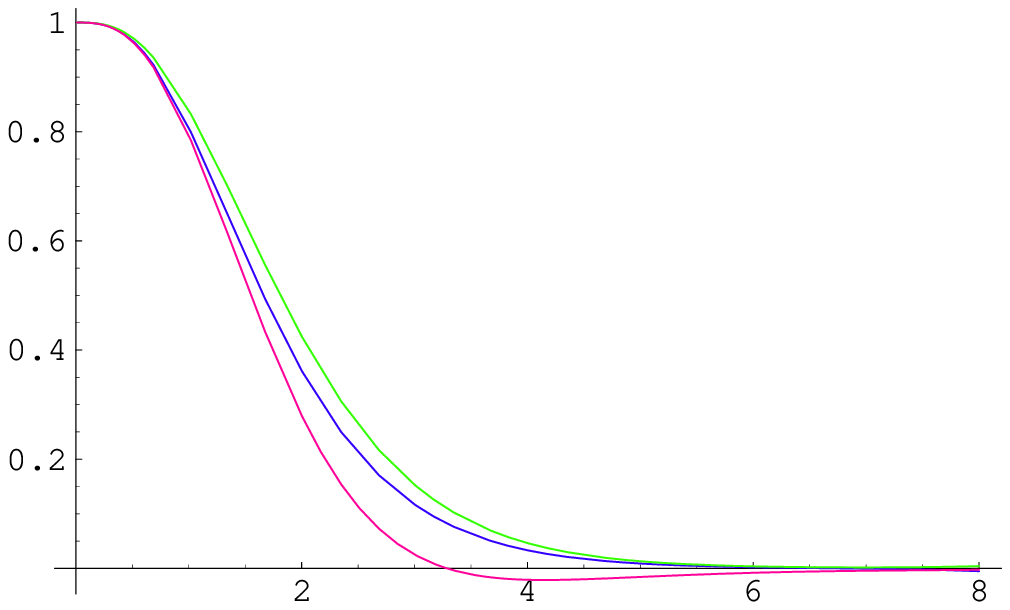}
\end{array}$
\end{center}
\caption{\footnotesize  f(r) (left):
$f_3(r)$ (center), $g(r)$ (right), 
at $\alpha=0$ (green), $\alpha=\frac{\pi}{2}$ (red), $\alpha=\pi$ (blue).
} \label{confronto2}
\end{figure}

Explicit numerical calculation can be and were performed  for the
 vortex profile
functions. The dependence on $\alpha$ is non-trivial.
Some of the profile functions at $\alpha=0$ (green),
$\alpha=\frac{\pi}{2}$ (red),
$\alpha=\pi$ (blue)
are plotted and compared in Fig. \ref{confronto1},\ref{confronto2}.
 The couplings are
chosen as
\beq 
e_0^2=1,\quad e_3^2=2\, .
\eeq
It seems that there is a small but non-trivial dependence on $\alpha$
This is evident, in particular, for $\kappa$, but also for $z_1,\, z_2$.

\subsection{Physical interpretation}
\label{physical}

To understand the explicit solution better it is instructive
to calculate the gauge invariant operator $ \tilde{\mathcal{F}}^a$.
It is possible to make a global SU(2)$_{C+F}$ rotation of the solution,
so that
  $\mathcal{F}^{(1,2)}$ averaged with respect to the 
azimuthal angle  $\varphi$ are zero.

 The following matrix  realizes this:
\beq 
\tilde{U}= 
 \exp\left(-i \alpha \frac{\tau_2}{2}\right),
 \eeq 
 acting on the field as 
\beq 
  \mathcal{F}^a \tau^a \rightarrow 
  \tilde{U}^{\dagger}\cdot \mathcal{F}^a \tau^a \cdot \tilde{U},
 \,\,\,Q \rightarrow  \tilde{U}^{\dagger}\cdot Q \cdot\tilde{U}.
 \eeq 

This gives us a ``minimal'' non-Abelian 2-string solution parametrized
by angle $\alpha$. To obtain the full moduli space of solutions
we have to apply the global SU(2)$_{C+F}$ rotation to the ``minimal''
solution.

The ``minimal'' solution has the form
{\small \beq Q =\sqrt{\xi} \kappa  \left(\begin{array}{cc}
\left(\cos^2  \frac{\alpha}{2} \right)   e^{2 i \varphi} z_1 z_2+
\left(\sin^2 \frac{\alpha}{2}\right) e^{ i \varphi} z_2
& -\frac{\sin \alpha}{2} (e^{ i \varphi}  z_1 - 1) \\[4mm]
-\frac{\sin \alpha}{2} (e^{2 i \varphi}  z_1 z_2 -  e^{ i \varphi} z_2 )
& \left(\cos^2 \frac{\alpha}{2} \right)  
+\left(\sin^2 \frac{\alpha}{2}\right) e^{ i \varphi} z_1 \\
\end{array}\right)\,,
\eeq }
and 
\beq 
\tilde{\mathcal{F}}^3 = A(\alpha,r),\,\,\,\,
 \tilde{\mathcal{F}}^1 =  (\cos\varphi) B(\alpha,r),\,\,\,\,
 \tilde{\mathcal{F}}^2 = -(\sin \varphi) B(\alpha,r)\,, 
 \label{bella}
 \eeq
where
{\small \beqn 
&& A(\alpha,z) =
 \left(\frac{g'}{r}+\frac{h (\cos \alpha-f_3)}{r^2}\right)
 \kappa^2 z_1 (z_2^2+1) (\sin \alpha)^2  
 \nonumber\\[3mm]
&& 
+ \left(\frac{f_3'}{r}- \frac{h(1-g)(\sin \alpha)^2}{r^2} \right)  \frac{\cos \alpha}{2}
 ( \kappa^2(z_1^2-1)  (z_2^2-1)
 + \kappa^2(z_1^2+1)  (z_2^2+1)\cos \alpha)
  \nonumber\\
\eeqn  }
  and 
 \beqn
&&  B(\alpha,z) =
  (\cos\alpha \sin \alpha)  \kappa^2 z_2
 \Big( 2 \left(\frac{g'}{r}+\frac{h(\cos \alpha-f_3)}{r^2}\right) z_1 
\nonumber\\[3mm] 
&& -\left(\frac{f_3'}{r}- \frac{h(1-g)(\sin \alpha)^2}{r^2} \right)(z_1^2+1) \Big)\, . \eeqn

 This solution at fixed $\alpha$ can be rotated
by applying an $SU(2)$ global color+flavor rotation.
 For generic $\alpha\neq 0, \pi$
  all  $SU(2)_{C+F}$ generators are
 broken by the vortex solution.
 The $\tau_{1,2}$ generators rotate the color flux direction which is
 independent of the cylindrical coordinate $\varphi$; the $\tau_3$ generator
 shifts a phase $\varphi$ in the arguments of 
 the sine and cosine functions in Eq. (\ref{bella}).
 The resulting moduli space is parametrized
 by the Euler angles, in complete analogy to the
 phase space of a cylindrical rotator in three-dimensional space.
 In  particular, for 
 $\alpha=0$ ($(2,0)$ vortex) we have $B=0$,
  and  for $\alpha=\pi$ ( $(1,1)$ vortex) we have $A=B=0$.
The behavior of the solution near these points   
 is discussed in Appendices; here we summarize our results at the 
 qualitative level.

\begin{figure}[h]
\begin{center}
\epsfxsize=4in \epsffile{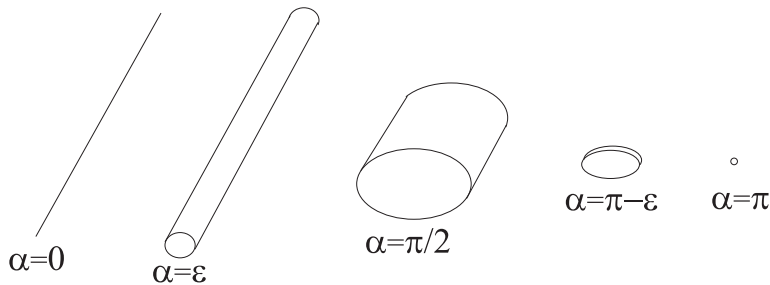}
\end{center}
\caption{\footnotesize The structure of the moduli space is very
similar to the phase space of  a cylindrical rotator  whose shape depends by
the parameter $\alpha$. At $\alpha=0$ one of the inertial moments is zero
(as for a stick with zero thickness); at $\alpha=\pi$ all the inertial moments
are zero.}
\label{pop-physics}
\end{figure}

Let us consider the solution as  a  function of  the angle $\alpha$
(see Fig.~\ref{axis}).
At $\alpha=0$ we have  the $(2,0)$ vortex; the action of the global $SU(2)$
is similar to the action of spatial rotation over a stick of
zero thickness and the moduli space is $S^2$. At 
small non-zero $\alpha$  the stick
acquires a thickness of order $\alpha$, and becomes in color space similar to
a cigarette. The moduli space  is now parametrized by three Euler angle
in color space and it is 3-dimensional.
Increasing $\alpha$ we can imagine that the cigarette becomes
shorter and fatter, becoming  a can.
At $\alpha=\pi-\epsilon$ the length becomes zero at the linear order 
in $\epsilon$;
on the other hand the diameter of our can is of order $\epsilon$. 
The configuration
in color space becomes similar to a coin with zero thickness: the 
moduli space
is still parametrized by three Euler angles.
At $\alpha=\pi$ our coin shrinks to a point and the action of global 
color-flavor
rotation is trivial.

\section{The 2-string moduli space}
\label{strms}

\subsection{Field theory perspective}
\label{ftpe}

The 2-vortex moduli space is a manifold with real dimension 8.
Two  coordinates correspond to a global translation and
 we factorize them from the other 6, which correspond to the non trivial
part of the moduli space:
\[ \mathcal{M}=\mathbb{C} \times \tilde{\mathcal{M}}. \]
In the limit of large relative distance between the two elementary vortices
 $\tilde{\mathcal{M}}$ has the following structure \cite{HashT} :
\beq
\label{msss} 
\tilde{\mathcal{M}} \approx
\frac{ \mathbb{C} \times  \mathbb{CP}^1 \times \mathbb{CP}^1}{\mathbb{Z}_2}, 
\eeq
where $\mathbb{C}$ corresponds to the relative distance of the two 
elementary vortices
and the two $ \mathbb{CP}^1$ factors 
stand for  the non-Abelian internal orientation
of the elementary vortices. The $\mathbb{Z}_2$ quotient acts 
on $(z,\,\,\vec{n}_1,\,\,\vec{n}_2)\in  \tilde{\mathcal{M}}$ as follows:
\beq 
\mathbb{Z}_2: \,\,\,\,\, z\rightarrow -z, \,\,\, \,\, \vec{n}_1 \rightarrow 
\vec{n}_2,
\,\,\,\,\, \vec{n}_2 \rightarrow \vec{n}_1\,.
\eeq

In the following we will discuss topology of the slice of the 
moduli space in which the relative distance of the elementary
vortices is zero. We denote this subspace by $\mathcal{T}$.
In the previous section we have found an explicit solution,
which  can be  parameterized   by
an $SU(2)\times SU(2)$ element $(U_1,U_2)$,
\beq
 Q =\sqrt{\xi} \kappa(r) U_1 \left(\begin{array}{cc}
 z_1(r) e^{i \varphi}& 0 \\[3mm]
0 & 1 \\
\end{array}\right)  U_1^{-1}\,
U_2  \left(\begin{array}{cc}
z_2(r) e^{i \varphi}& 0 \\[3mm]
0 & 1 \\
\end{array}\right) U_2^{-1}\,,
\eeq 
where
\beq 
U_1 \tau_3 U_1^{\dagger}=n_1^a \tau_a, \qquad U_2 \tau_3 U_2^{\dagger}=n_2^a \tau_a\,.
\eeq
The functions $\kappa,z_1,z_2$ depend on the relative angle $\alpha$ between 
$\vec{n}_1$ and $\vec{n}_2$ in a non-trivial way.
Taking $U_1=U_2=U$ the usual global orientation
zero modes are obtained.
Each of the $SU(2)$ subgroups is broken down locally
to $U(1)$. However the situation is different globally:
for example, taking  $\vec{n}_1=-\vec{n}_2=\vec{n}$
we find just a point in the moduli space (the $(1,1)$ vortex) rather then
a 2-dimensional submanifold.
So $\mathcal{T}$ is not 
$\mathbb{CP}^1 \times \mathbb{CP}^1$ as one could naively expect.

Let us consider  topology of different slices at constant $\alpha$.
At $\alpha=0$, the moduli space is given by
 \[ \mathcal{T}_{\alpha=0}=SU(2)/U(1)=\mathbb{CP}^1=S^2. \]
At $0<\alpha<\pi$ the moduli space is given by the quotient 
\beq
 \mathcal{T}_{0<\alpha<\pi}=SU(2)/\mathbb{Z}_2=\mathbb{RP}^3=S^3/\mathbb{Z}_2\,,
 \label{sss} 
\eeq
because the global rotations from  the center of
$SU(2)$ have  trivial effect on the solution. 
At $\alpha=\pi$, the moduli space is just a point
rather then  a manifold.
If it were a manifold, then a submanifold of constant
 small $\alpha$ would be topologically equivalent to $S^3$,
but we know that it is $\mathbb{RP}^3$, which differs from $S^3$
by a $Z_2$ quotient. We conclude that at $\alpha=0$
there is  a conical singularity;
this is similar to the singularity in the 1-instanton
moduli space for the zero-size instanton. For a dedicated discussion 
of the occurrence of the  $\mathbb{Z}_2$ factor
in Eq.~(\ref{sss}) see Sect.~\ref{twovortexbrane}.

Topology of $\mathcal{T}$ is equivalent to
a discrete quotient of $\mathbb{CP}^2$.
To make it clear we use the following parametrization of 
$\mathbb{CP}^2$:
\beq
\label{cp2}
\vec{m}=(m_1,m_2,m_3),
\eeq
where $m_i$ ($i=1,2,3$) are complex variables subject to the
constraint
\beq
\label{normone}
|m_1|^2 + |m_2|^2 +|m_3|^2 =1
\eeq
and identification
\beq
\label{id}
\vec{m} \sim e^{i\delta}\vec{m}\, .
\eeq
Complex vector $\vec{m}$ has six real variables. Condition (\ref{normone})
and identification (\ref{id}) reduce this number to four, which is 
the dimension of $\mathbb{CP}^2$. 

The variable $| m_1|$ plays a role
of $\sin{\alpha/2}$ for our solution. At $\alpha =0$ (i.e. (2,0) string)
the vector $\vec{m}$ has only two components and parametrizes $\mathbb{CP}^1$
manifold which is a moduli space of the (2,0) string indeed. At $\alpha=\pi$
$$m_2=m_3=0$$ and the space described by the vector $\vec{m}$ shrinks to a point,
just like the moduli space of the (1,1) string. At intermediate $\alpha$,
$$0<\alpha<\pi$$ the  vector $\vec{m}$
produces $ SU(2)=S^3$ submanifolds.
We conclude that   
  topology of the 2-string moduli space $\mathcal{T}$
is given by the following quotient:
\[ \mathcal{T}=\mathbb{CP}^2/\mathbb{Z}_2,\]
where $Z_2$ acts as
\beq 
(m_1,m_2,m_3)\rightarrow (m_1,-m_2,-m_3).
\eeq
This $Z_2$ subgroup acts trivially at $\alpha=\pi$
(where $\vec{m}=(1,0,0)$) and at $\alpha=0$
(where $\vec{m}=(0, m_2,m_3)$) because of the identification 
(\ref{id}).
The sections at constant $\alpha$ with $0<\alpha<\pi$
have the topology of $\mathbb{RP}^3=S^3/Z_2$.
Near $\alpha=\pi$  there is  a conical singularity.

When one chooses a particular {\em ansatz},
generally speaking, one is not guaranteed that in this given
{\em ansatz} all moduli space of the solitonic object at hand is
covered. In principle, it could happen that an {\em ansatz}
containing an appropriate number of collective coordinates 
is  still not  general enough in order to describe in full the family of solutions. 
We would like to argue that
this is not the case here ---
 we do cover all the moduli space 
 of two {\em coincident} vortices. Our {\em ansatz} has the
right number of  collective coordinates; it is not singular anywhere on
the moduli space. Moreover,  we expect that $\mathcal{T}$ is a 
topological space with just a single connected component.
Finally, let us stress that the $Z_2$ quotient (a subtle point
of the construction) appears as a consequence of the $SU(2)$
global rotations  rather than as a specific feature of 
the particular form of our {\em ansatz}.
As a nontrivial check, we will show in Sect.~\ref{twovortexbrane}, with satisfaction,
that the result agrees with one from of the brane construction.

The effective $(1+1)$-dimensional theory on the string world sheet
is a sigma-model  determined by the metric
on the vortex moduli space. 
We know from $SU(2)_{C+F}$ symmetry arguments that
the metric on $\mathcal{T}$ has the form of a cylindrical
 rotator with an extra parameter $\alpha$,
\beqn 
w d \alpha^2 &+&\frac{1}{2} \left[ I_{xy}\,  d\theta^2 +\frac{(I_z+I_{xy})+(I_z-I_{xy}) \cos 2 \theta }{2} \, d\phi^2\right.
\nonumber\\[3mm]
&+&
I_z \, d\psi^2 + 2 I_z \, \cos \theta  \, \, d \phi\,\,  d \psi\,   \Big]\,,
\eeqn
where $\theta$, $\phi$ and $\psi$ are Euler angles while $\alpha$
ia an extra parameter,
 $0<\alpha<\pi$. Explicit determination
of the functions $w(\alpha)$, $J_{xy}(\alpha)$ and $J_z(\alpha)$
remains  an open problem.

\subsection{The 2-vortex in the brane construction}
\label{twovortexbrane}

In Ref. \cite{HT} and \cite{HashT} a construction for topology
of the 2-vortex moduli space was proposed within the Hanany-Witten approach.
In these papers it is shown that the moduli space of $k$ vortices 
in the $U(N)$ theory with $N_f=N$ flavor hypermultiplets
is a K\"{a}hler  manifold with real dimension
$2k N_c$ that we will denote as $\mathcal{H}_{k,N}$.
The K\"{a}hler  manifold  $\mathcal{H}_{k,N}$
is built as follows.

 Let us start with a $k \times k$ complex matrix $Z$
and a $k \times N$ complex matrix $\Psi$, with the constraint
\beq 
[Z,Z^\dagger]+\Psi \Psi^\dagger=1 \label{hth}\,,
 \eeq
 where $1$ is the identity matrix.
The space  $\mathcal{H}_{k,N}$ is defined as the quotient of the 
solution of this constraint divided by the $U(k)$ action,
\beq 
Z \rightarrow U Z U^\dagger, \qquad
\Psi \rightarrow U \Psi\,. 
\label{creedence} 
\eeq 
The manifold  $\mathcal{H}_{k,N}$ has the symmetry $SU(N)\times U(1)$,
\beqn
&& SU(N): \,\, \Psi \rightarrow \Psi V, \,\,\,V \in SU(2), \nonumber\\[3mm]
&& U(1): \,\, Z \rightarrow e^{i \alpha }Z\,. 
\eeqn
In this formalism  the action of the $SU(N)$ group is physically identified with
the $SU(N)_{C+F}$ while that of  the $U(1)$ with the rotational symmetry
of the plane.

In the case of 2-strings in the $N_f=N=2$ gauge theory
both $Z$ and $\Psi$ are $2 \times 2$ matrices. Requiring 
$\Tr Z=0$ we project out the trivial center-of-mass motion. 
The action of Eq. (\ref{creedence}) can be used to transform $Z$ in the
upper-triangular form,
\beq Z= \left(\begin{array}{cc}
 z & \omega \\
0 & -z  \\
\end{array}\right), \qquad \Psi= \left(\begin{array}{cc}
 a_1 & a_2 \\
b_1 & b_2  \\
\end{array}\right). \eeq
The coordinate $z$ represents the relative positions of the strings;
the other entries of the matrices have less intuitive interpretation.
This does not completely fix the $U(2)$ quotient; a remaining 
$U(1)_1 \times U(1)_2 \times \mathbb{Z}_2$ has to be fixed, namely,
\beqn 
&& U(1)_1: \,\, U=\left(\begin{array}{cc}
 e^{i \phi} & 0 \\
0 & 1  \\
\end{array}\right), \,\,\, 
U(1)_2: U=\left(\begin{array}{cc}
 1 & 0 \\
0 & e^{i \phi}  \\
\end{array}\right),
\nonumber\\[4mm]
&& \mathbb{Z}_2: \,\, U= \frac{-1}{\sqrt{1+|2z/\omega|^2}} 
\left(\begin{array}{cc}
 -1 & (2z/\omega)^* \\[2mm]
(2z/\omega) & 1  \\
\end{array}\right).
\eeqn
We have the following charges with respect $(U(1)_1,U(1)_2)$:
 $$a_i \rightarrow (1,0),\quad b_i \rightarrow (0,1),\quad
\omega \rightarrow (1,-1),\quad z \rightarrow (0,0)\,.$$
The constraints in Eq. (\ref{hth}) read
\beq   |a_1|^2+|a_2|^2+|\omega|^2=1, \,\,\,
 |b_1|^2+|b_2|^2-|\omega|^2=1, \,\,\, \sum a_i b_i^*=2z^* \omega. \label{ciccio} \eeq

If we put $z=0$ we can recover topology of $\mathcal{T}$.
Let us consider, following Ref. \cite{HashT}, slices at constant $\omega$.
At $\omega=0$ a point is found which is the $(1,1)$ vortex (note that the entries of the matrix $Z$ all vanish and   all   $U(2)$ quotient has to be fixed for the matrix $\Psi$).
At $|\omega|=1$ a copy of $\mathbb{CP}^1$ is found which  is 
the $(2,0)$ vortex and its color-flavor rotated configurations. (This is because $a_i=0$ and  $b_i$  define a $\mathbb{CP}^1$ modulo the $U(1)_2$ action). 

The slices at $0<|\omega|<1$ are slightly  more complex.
Let us consider them in detail. Let us define $U(1)_A$ as 
$U(1)_1+U(1)_2$ and $U(1)_B$ as $U(1)_1-U(1)_2$. 
We have the following charges with respect to $(U(1)_A,U(1)_B)$:
 $$
 a_i \rightarrow (1,1),\quad b_i \rightarrow (1,-1),\quad
\omega \rightarrow (0,2)\,.
$$
The most general solution to the constraints in Eq. (\ref{ciccio}) is
\begin{eqnarray}
 a 
 &=&
  \left(1-|\omega|^2\right)  e^{i \sigma} \left(\cos \theta, \,\sin \theta e^{i \phi}
 \right) ,
 \nonumber\\[2mm]
 b
 &=&
  \left(1+|\omega|^2 \right)  e^{i \eta}  \left(-\sin \theta,\, \cos \theta e^{-i \phi} \right) ,
 \nonumber\\[2mm]
\omega
 &=&
e^{i \gamma}\, |\omega|\, . 
\label{ssss}
\end{eqnarray}
The quotient $U(1)_B$ just gauges away the phase $\gamma$ so that
effectively $\gamma =0$,
with a redefinition of $\sigma$ and $\eta$. Using $U(1)_A$  at this
point we can bring the solution to the following
form (where $\delta=(\sigma-\eta)/2$):
\begin{eqnarray}
 a
 &=&
   \left(1-|\omega|^2\right)  e^{i \delta}   \left(\cos \theta, \,\sin \theta \, e ^{i \phi}\right) ,
  \nonumber\\[2mm]  
 b
 &=&
 \left(1+|\omega|^2 \right) e^{-i \delta}  \left(-\sin \theta, \cos \theta \, e^{-i \phi} \right),
   \nonumber\\[2mm]  
\omega
 &=&
 |\omega| \,.
\label{fulcro}
\end{eqnarray}
The three angles $(\theta,\delta,\phi)$ parameterize an $S^3$ inside $\mathbb{C}^4$.  We have to be   careful, however,  because we still have to perform a quotient  in order to find the moduli space. Namely, in
 $S^3$ we have to identify the
opposite points as
\beq
 (a_i,b_i,|\omega|) \rightarrow (-a_i,-b_i,|\omega|) 
 \eeq
because if we shift $\delta$ by $\pi$ we have that both $a_i,b_i$ get a $-1$ phase which is exactly a $\pi$ rotation by $U(1)_A$. This
special rotation keeps the solution in the form of Eq. (\ref{fulcro}), and,
therefore,  we have to take account of  this
special rotation  ``by hand." 
In other words, when
we put the solutions of the constraints in the form  (\ref{fulcro}), we fixed {\sl  almost} all  gauge freedom, with the exception of a $\mathbb{Z}_2$ subgroup generated by a $\pi$ rotation by $U(1)_A$.

We conclude that our solitonic solution is consistent
with the  brane technique-based results.
The $\alpha=\pi$ section in the field-theory approach corresponds to $\omega=0$ in the brane construction (the $(1,1)$ vortex); $\alpha=0$ corresponds to $\omega=1$ (the $(2,0)$ vortex moduli space). Sections at intermediate $\alpha$ and $\omega$ are in 
the both cases
$S^3/\mathbb{Z}_2$, and at the end the both   approaches 
give $\mathcal{T}=\mathbb{CP}^2/\mathbb{Z}_2$.

\section{Confined monopoles}
\label{seven}

If the Fayet-Iliopoulos term $\xi$ vanishes, the
squark condensate vanishes too, and the
theory is in Coulomb phase. Then there exists the t'Hooft-Polyakov
monopole, and its magnetic flux is unconfined.
When a non-vanishing $\xi$ is introduced, the squarks develop
a VEV, and the theory is in the Higgs phase. The monopole flux
is confined.  In our theory there is 
a stable  configuration for  the monopole 
confined by two strings oriented in opposite directions.
In this configuration the monopole flux is carried by
 two elementary flux tubes
 (see \cite{tong-monopolo,SY-vortici,HT2,012}).
This monopole can be interpreted
as the junction of two different magnetic strings.

If $\Lambda \ll  \Delta m \ll  \xi^{1/2}$ the quasi classical treatment is
 reliable.
We find that the monopole is a classical soliton
which is the junction of the
$(1,0)$ and  $(0,1)$ strings.
 The composite monopole+vortex object is $1/4$ BPS;
the energy is given by the BPS bound:
\beq  \int \mathcal{H} d^3 x =\int \Tr \left[ 
\xi B_z - \frac{1}{e_3^2} \partial_\alpha (a\cdot B_\alpha) \right] d^3 x =
\int T_v dz  + M_{\rm mon}. \eeq
where
\beq T_v=2 \pi \xi, \,\,\,\, M_{\rm mon}= \frac{2 \pi (m_1-m_2)}{e_3^2}. \eeq
The effective world sheet
description is given by an $\mathcal{N}=2$ $\mathbb{CP}^1$
 sigma model with a large twisted mass term $\mu = \Delta m$,
which has two classical vacua (see \cite{dorey1,dorey2,SY-vortici,HT2}).

In the limit $ \Delta m \ll \Lambda\ll \xi^{1/2} $ 
the situation is more subtle; the monopole
is not a classical object. The vortex world-sheet theory
is an $\mathcal{N}=2$ $\mathbb{CP}^1$ sigma model.
Classically this model has an infinite number of 
vacua parametrized by
 points of $\mathbb{CP}^1$ and there are Goldstone states.
In  quantum theory
  due to non-perturbative 
effects all states become massive.
The  theory  has two quantum vacua,
 as can be shown by Witten-index arguments. These two vacua
correspond to two quantum non-Abelian strings.
The monopole can be interpreted as a kink between these
two vacua;
 the monopole mass is given  by the mass of the kink
in the $1+1$ dimensional sigma model \cite{SY-vortici}
\beq 
M_{\rm mon}=\frac{2}{\pi} \Lambda_{\mathbb{CP}^1},
\eeq
where $\Lambda_{\mathbb{CP}^1}= \Lambda_{QCD}$.
In both the limits we have 
 two physical string states and a confined monopole
which can be interpreted as the junction between these strings.

Let us consider what happens for the case of the composite
2-vortex. If $ \Delta m \gg  \Lambda$ we have    Abelian vortices with
the same tension, the $(2,0)$, $(0,2)$ and $(1,1)$ vortices.
There are two possible kinds of confined monopoles:
the one between the $(2,0)$ and the $(1,1)$ vortices
and the one between the $(2,0)$ and the $(0,2)$.
If we calculate the monopoles masses using the central charge,
we find that:
\beq M_{(2,0)\rightarrow(1,1)}=
\frac{2 \pi (m_1-m_2)}{e_3^2}=\frac{M_{(2,0)\rightarrow(0,2)}}{2}\eeq
We can think the $(2,0)\rightarrow(0,2)$ kink as the composite state
of the $(2,0)\rightarrow(1,1)$ and the $(1,1)\rightarrow(0,2)$
kinks; it is reasonable that there is no net force between the
two elementary kinks because the energy of the bound state is equal
to the sum of masses of  two elementary kinks.

\begin{figure}[h]
\begin{center}
\epsfxsize=3.5in \epsffile{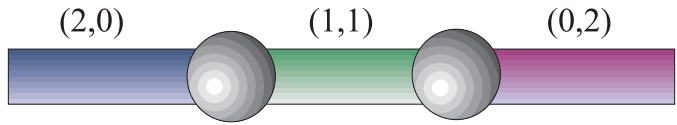}
\end{center}
\caption{\footnotesize Elementary BPS confined monopoles connecting
 the $(2,0)$ to the $(1,1)$ vortex and the $(1,1)$ to the $(0,2)$ vortex. 
 The mass of a BPS monopole connecting the $(2,0)$ to the $(0,2)$ vortex
 is exactly the double of an elementary monopole. We can conjecture that
 the length of the intermediate layer of $(1,1)$ vortex is a modulus of
 the composite soliton.}
\label{oberon}
\end{figure}

When we go to the limit $ \Delta m \ll  \Lambda$ 
 the situation becomes rather complicated.
 Even if we neglect for simplicity the 
the coordinate corresponding to the relative distance of the elementary
 vortices, 
 the physics is described by a sigma model
with target space $\mathcal{T}=\mathbb{CP}^2/\mathbb{Z}_2$ 
(a space that is not even
a manifold due to a conical singularity) and with a quite complicated metric.
However, in analogy with  the 1-vortex  case it is reasonable 
to think that the
spectrum of BPS states in 2-dimensional world sheet model coincides
with the monopole/dyon spectrum of the 4-dimensional bulk theory
on the Coulomb branch because it cannot depend on FI parameter
$\xi$ \cite{SY-vortici,HT2}.
The latter spectrum is given by exact Seiberg-Witten solution
\cite{SW1,SW2}.

\section{Conclusions}
\label{eight}

In this paper we considered a composite non-Abelian vortex 
with winding number 2 in $\mathcal{N}=2$ supersymmetric theory 
with gauge group $U(2)$.

The explicit BPS solution  of first order equations  has been found
in the case when two component elementary vortices are parallel
and coincident in the space.

The internal moduli space $\mathcal{T}$ has the topology
$\mathbb{CP}^2/\mathbb{Z}_2$; there is a conical singularity
near the $(1,1)$ vortex.
The computation of the metric for the effective sigma model on
$\mathcal{T}$ still remains   an open question.
However, perturbing the system with a $\Delta m$, it is possible
to guess the number of vacua and the spectrum of kinks
in the $1+1$ dimensional effective description.

\section*{Acknowledgments}

We are grateful to Jarah Evslin, Andrei Losev and Arkady Vainshtein for useful
discussions.
The work  of R.A. and M.S. is
supported in part by DOE grant DE-FG02-94ER408.
The work of A.Y. is supported by Theoretical Physics Institute
at the University of Minnesota.

\section*{Note Added}

After this work was finished, a paper by Eto et al. was submitted \cite{new}.
Eto et al. extended the analysis of Ref.~\cite{HashT}
and thus completed a construction allowing one to introduce the
full number $2Nk$ of (real) collective coordinates in the generic
$k$-string BPS solution.
The problem of introduction and actual calculation of the
profile functions was not addressed. 
Our result is complementary albeit not generic.
One can show that our ansatz, being cast in the form suggested in
\cite{new}, reduces to
\beq
H_0 (z) =\left(
\begin{array}{ll}
-\cos\frac{\alpha}{2}\, z^2 & \sin \frac{\alpha}{2}\, z\\[3mm]
-\sin \frac{\alpha}{2}\, z & -\cos\frac{\alpha}{2}
\end{array}
\right)
\label{addl}
\eeq
in  the gauge discusssed in Sect.~\ref{another}, modulo global $SU(2)$
rotations (which introduce other three collective coordinates).
The determinant of the matrix above is $z^2$,
with a degenerate zero at the origin,  which is  a signal,
in the language of Ref.~\cite{new}, of the coincidence of the positions
of two constituents of the 2-string under consideration.
It seems very plausible that applying the general method of 
\cite{new} one can extend our ansatz to include
two missing collective coordinates responsible for the relative separation
 of two constituents of the 2-string in the perpendicular plane.

\section*{Erratum to \\
{\underline{\sl Composite Non-Abelian Flux Tubes in $\mathcal{N}=2$ SQCD}},\\
by R.~Auzzi, M.~Shifman and A.~Yung,\\
Phys.\ Rev.\  D {\bf 73}, 105012 (2006)\\
September 14, 2007 }

Section 5 of the above paper presents an ansatz for the two-string solution.
This ansatz turns out to be too restrictive. Some  details
must be changed.  Basic results and conclusions of the paper remain intact,
in particular our main conclusion on
topology of the moduli space of the composite string.

The corrections to be introduced in the ansatz presented in Sect.~5 of the 
above paper are as follows.

The rotation discussed in Sect.~5.4 is not a color rotation,
but rather a flavor one. Therefore, the expression for the gauge fields in Eq.~(44) is wrong.
Equation~(32) is still the correct, also after the flavor rotation.
After the flavor rotation the ansatz for $Q$ used in the paper is (see Eq. (43))
 \beq 
 Q = \sqrt{\xi} \kappa \left(\begin{array}{cc}
-\cos \frac{\alpha}{2}  e^{ 2 i \varphi}  z_1 z_2
& \sin \frac{\alpha}{2}  e^{  i \varphi}  z_1  \\[3mm]
- \sin \frac{\alpha}{2}  e^{  i \varphi}  z_2 
& -\cos \frac{\alpha}{2}  \\
\end{array}\right). \label{old}
\eeq 
This expression is not sufficiently general. 
It must be replaced by
 \beq 
 Q =  \left(\begin{array}{cc}
-\cos \frac{\alpha}{2}  e^{ 2 i \varphi}  \kappa_1
& \sin \frac{\alpha}{2}  e^{  i \varphi} \kappa_2  \\[3mm]
- \sin \frac{\alpha}{2}  e^{  i \varphi}  \kappa_3 
& -\cos \frac{\alpha}{2} \kappa_4  \\
\end{array}\right), \label{new}
\eeq  
where $\kappa_{1,2,3,4}$ are functions of $r$
which tend to $\sqrt{\xi}$ at $r \rightarrow \infty$,
while at small $r$ 
\[\kappa_1(r)\to  \mathcal{O}(r^2),\,\,\, \kappa_2(r)\rightarrow \mathcal{O}(r), \,\,\,
\kappa_3(r)\rightarrow \mathcal{O}(r), \,\,\, \kappa_4(r)\rightarrow 
\mathcal{O}(1)\,. \] 
If  the ansatz of Eq.~(43) were correct, we would have
\beq \kappa_1 \kappa_4 = \kappa_2 \kappa_3. \eeq
In fact, this is not the case, as follows from numerical calculations with
the new ansatz \cite{aev}.  

Moreover, the function $h(r)$ used in the ansatz for the gauge field,
Eq.~(32), vanishes. In other words, the gauge field has no radial component.
The corrected form for the gauge field can be found 
replacing $h=0$ in Eq.~(32) by
\bea 
&& A^0_{(i)} = -\frac{\epsilon_{ij} x_j}{r^2}  (2-f), \,\,\,
 A^3_{(i)} = -\frac{\epsilon_{ij} x_j}{r^2}  ((1+\cos \alpha )-f_3), \label{ans2}  
\nonumber\\[3mm]
&&A^1_{(i)} = -\frac{\epsilon_{ij} x_j}{r^2} (\sin \alpha) (\cos\varphi)  (1-g), \nonumber\\[3mm]
&&A^2_{(i)} = +\frac{\epsilon_{ij} x_j}{r^2} (\sin \alpha) (\sin \varphi)  (1-g).
\eea
The expressions in Eq.~(35) for the field strength tensor are still correct,
with $h=0$.
In terms of the new profile functions, the BPS
equations (37) are replaced by the following system
of seven first-order equations:
\bea 
&&\frac{f'}{r}=\frac{e_0^2}{2} \left\{ \left( \cos \frac{\alpha}{2}\right)^2
(\kappa_1^2+\kappa_4^2)+ \left( \sin \frac{\alpha}{2}\right)^2
(\kappa_2^2+\kappa_3^2)-2  \xi \right\},
 \nonumber\\[3mm]
&&\frac{f_3'}{r}=\frac{e_3^2}{2} \left\{ \left( \cos \frac{\alpha}{2}\right)^2
(\kappa_1^2-\kappa_4^2)+ \left( \sin \frac{\alpha}{2}\right)^2
(\kappa_2^2-\kappa_3^2)\right\},
\nonumber\\[3mm]
&& \frac{g'}{r}=\frac{e_3^2}{2} \left\{ \kappa_1 \kappa_3-\kappa_2 \kappa_4 \right\},
 \nonumber\\[3mm]
&&\kappa_1'=  \frac{g-1}{r}  \sin ^2\left(\frac{\alpha }{2}\right) \kappa_3
+\frac{1-\cos (\alpha )+f+f_3}{2 r} \kappa_1,
\nonumber\\[3mm]
&&\kappa _2'=-\frac{g-1}{r}  \cos ^2\left(\frac{\alpha }{2}\right)\kappa _4 -
\frac{1+\cos (\alpha )-f-f_3}{2 r} \kappa _2, 
\nonumber\\[3mm]
&&\kappa _3'=\frac{g-1}{r}  \cos ^2\left(\frac{\alpha }{2}\right)\kappa _1 +
\frac{1+\cos (\alpha )+f-f_3}{2 r} \kappa _3, 
\nonumber\\[3mm]
&& \kappa _4'=-\frac{g-1}{r}  \sin ^2\left(\frac{\alpha }{2}\right)\kappa _2 -
\frac{1-\cos (\alpha )-f+f_3}{2 r} \kappa _4. 
\eea

The conclusions of Appendices A and B
regarding the $\alpha=0,\pi$ limits are intact.
The match with the profile functions of the Abelian
string (see Eq.~(11)) is simpler.
At $\alpha=0$, the vortex can be described by
Eq.~(11) of the paper, with $(p,k)=(2,0)$ and $\phi_1=\kappa_1$,
$\phi_2=\kappa_4$. The profile functions $\kappa_2,\kappa_3$ disappear
from the ansatz at $\alpha=0$, but, on the other hand, tend to 
well-defined functions in the limit 
$\alpha \rightarrow 0$.
At $\alpha=\pi$, the vortex becomes Abelian and can be described 
(after diagonalization) by
Eq.~(11) of the paper, with $(p,k)=(1,1)$ and $\phi_1=\kappa_2$,
$\phi_2=\kappa_3$. The profile functions $\kappa_1,\kappa_4$ disappear
from the ansatz at $\alpha=\pi$, but, on the other hand, tend to 
well-defined functions in the limit 
$\alpha \rightarrow \pi$.

\begin{figure}[h]
\begin{center}
$\begin{array}{c@{\hspace{.2in}}c} \epsfxsize=2.2in
\epsffile{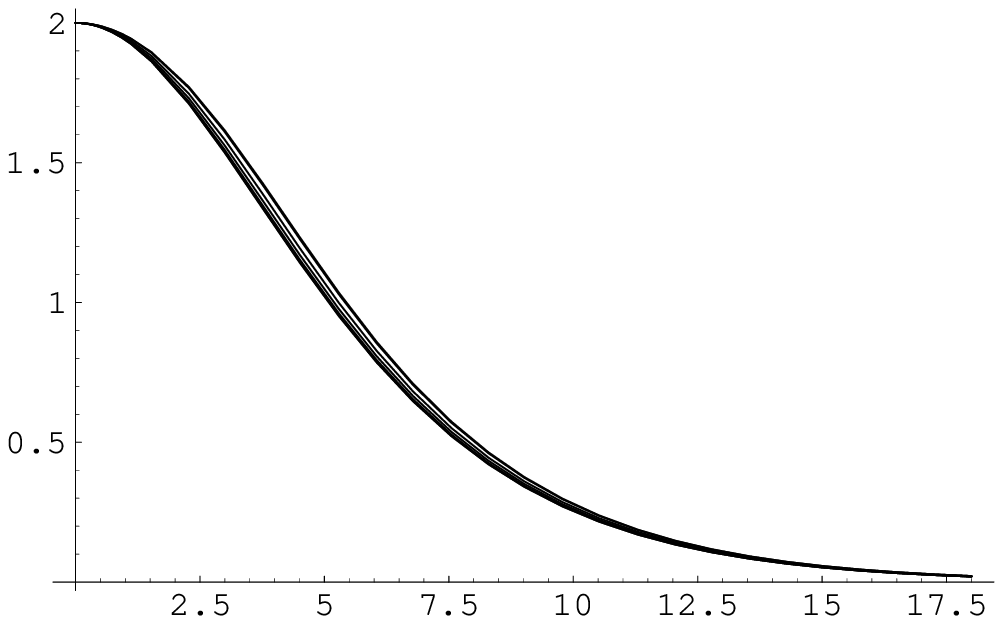} &
    \epsfxsize=2.2in
    \epsffile{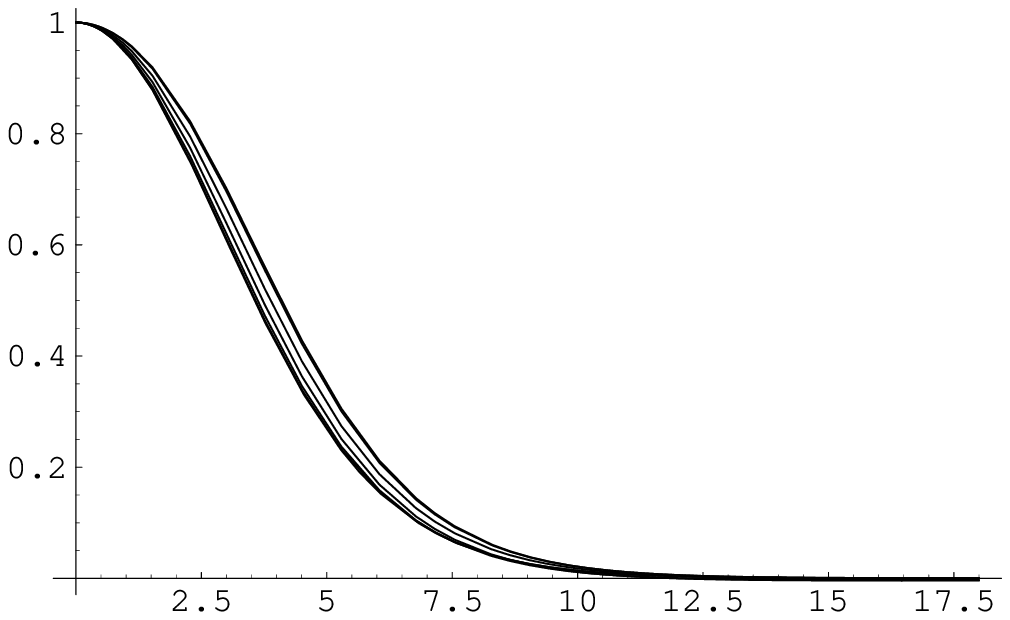} 
\end{array}$
\end{center}
\caption{\footnotesize Profile functions  $f_0$ (left) and  $f_3/(1+\cos(\alpha))$ (right).}
\label{fig1}
\end{figure}

\begin{figure}[h]
\epsfxsize=2.2in
\centerline{\epsfbox{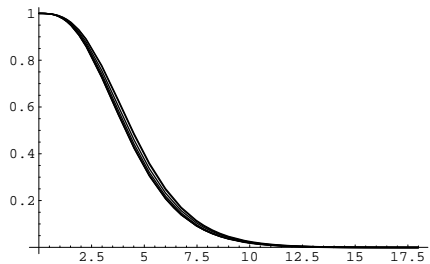}}
\caption{\footnotesize Profile function  $g$.    }
\label{fig2}
\end{figure}

\begin{figure}[h]
\begin{center}
$\begin{array}{c@{\hspace{.2in}}c} \epsfxsize=2.2in
\epsffile{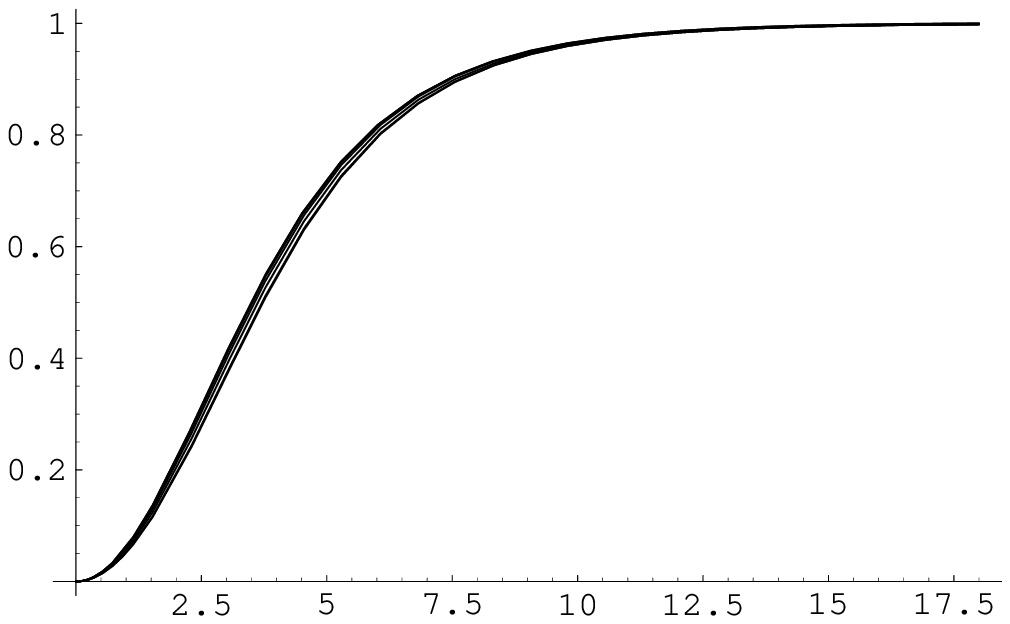} &
    \epsfxsize=2.2in
    \epsffile{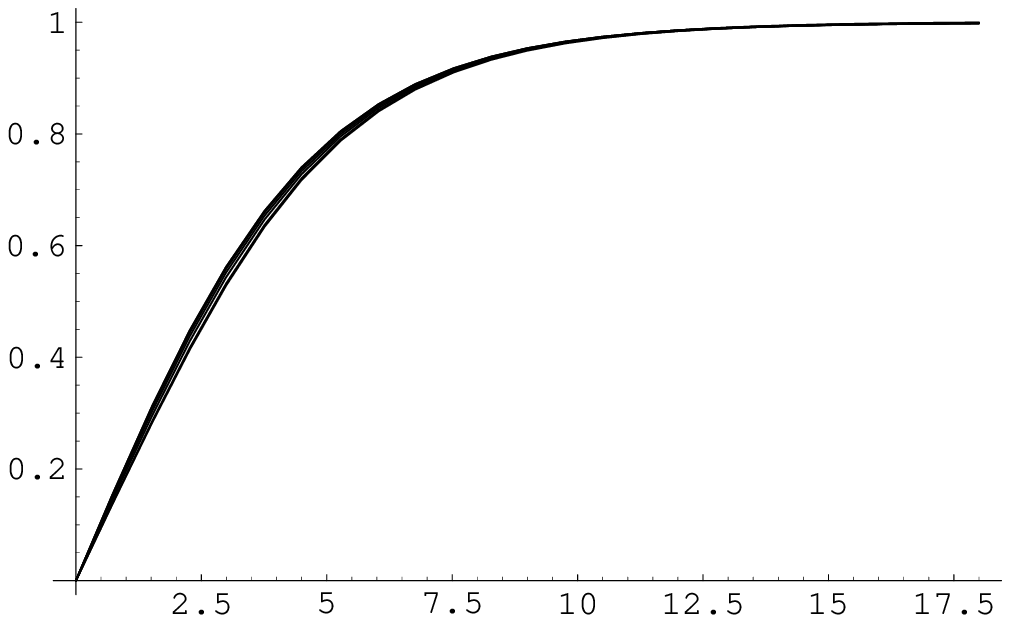} 
\end{array}$
\end{center}
\caption{\footnotesize Profile functions  $\kappa_1$ (left) and $\kappa_2$ (right).}
\label{fig3}
\end{figure}

\begin{figure}[h]
\begin{center}
$\begin{array}{c@{\hspace{.2in}}c} \epsfxsize=2.2in
\epsffile{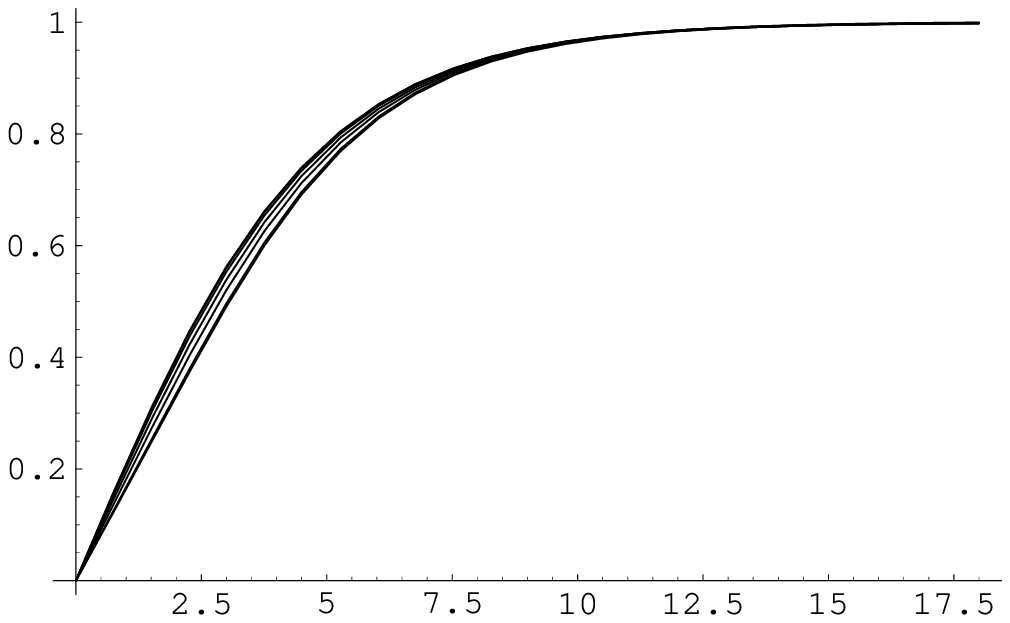} &
    \epsfxsize=2.2in
    \epsffile{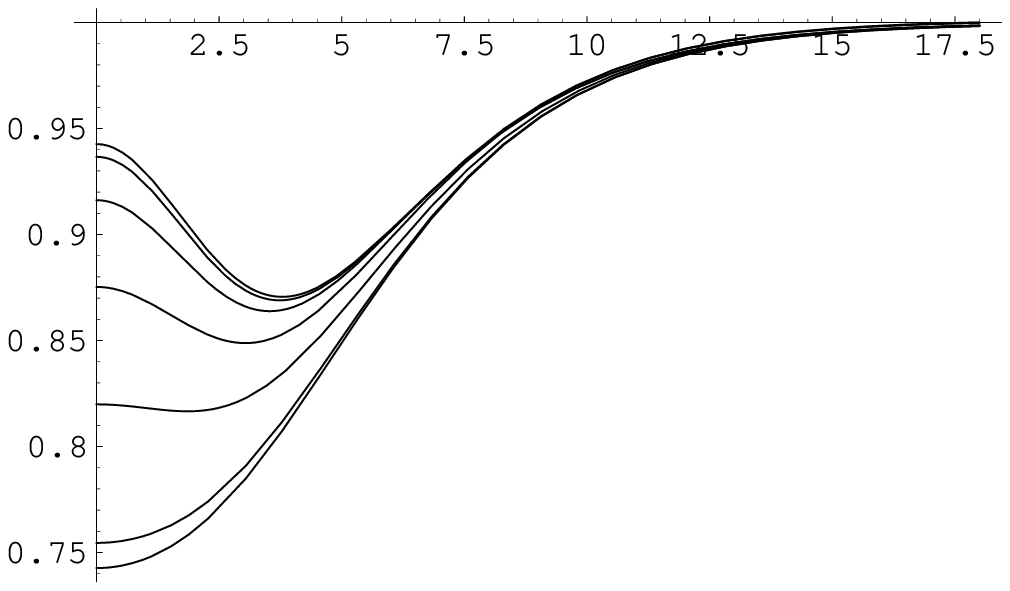} 
\end{array}$
\end{center}
\caption{\footnotesize Profile functions  $\kappa_3$ (left) and $\kappa_4$ (right).
For $\kappa_4$ the lowest  curve 
corresponds to $\alpha \rightarrow 0$, the highest one to $\alpha \rightarrow \pi$.}
\label{fig4}
\end{figure}

Results of numerical calculations of the profile functions within the new ansatz
at various values of $\alpha$ for $\xi=1,\,\,\, e_0=1/4,$ and $e_3=1/2$ are shown
in Figs.~\ref{fig1}, \ref{fig2}, \ref{fig3}, and \ref{fig4}.
The $\alpha$ dependence  is weak but nontrivial.
The profile function with the strongest $\alpha$ dependence is $\kappa_4$
(see Fig.~\ref{fig4} on the left).
The values used for $\alpha$ are $(\pi/20)$, $(\pi/5)$, $(\pi/2)$, $(2 \pi /3) $,
$(4 \pi/5)$, $(9 \pi /10)$ and $(19 \pi /20)$.

\newpage

\appendix

\section*{Appendix A: Zero modes for the \boldmath{$(1,1)$} vortex}
\renewcommand{\theequation}{A.\arabic{equation}}
\setcounter{equation}{0}

Let us consider a small perturbation around the
$(1,1)$ vortex at $\alpha=\pi$; let us
write $\alpha=\pi + \tilde{\alpha}$. 
All  profile functions are non-trivial 
functions of $\tilde{\alpha}$.
We will calculate the corrections to the profile functions 
at the first non-trivial order in $\tilde{\alpha}$.
Then we  substitute our solution into
 the action, see Eq. (\ref{azione}) and 
check that the linear and the quadratic corrections 
in $\tilde{\alpha}$ to the
tension are zero. 
To this order it is consistent to consider $\kappa,z_1,z_2$
as constants in $\alpha$.
We calculate  profile functions
$h$ and $g$ to  $\mathcal{O}(1)$ order in $\tilde{\alpha}$,
however 
$f$ should be be calculated with higher accuracy, namely to the 
 order $\mathcal{O}(\tilde{\alpha}^2)$.
Notice that at this order it is consistent to take $z_1=z_2=z$ and
as a consequence
$f_3=\mathcal{O}(\tilde{\alpha}^2)$
(this follows from  the BPS equations 
\beqn
&& \frac{f'_3}{r}-\tilde{\alpha}^2\frac{h(1-g)}{r^2}=\frac{e_3^2}{2}
( \kappa^2 (z_1^2-z_2^2))\,,
\nonumber\\[3mm]
&&\frac{f_3}{r}= \frac{z_1'}{z_1} -\frac{z'_2}{z_2} 
(1+\mathcal{O}(\tilde{\alpha^2}))
\label{a1}
\eeqn
combined with
 the boundary condition 
$f_3(r\rightarrow 0)=\mathcal{O}(\tilde{\alpha}^2)$);
so we will not need to compute $f_3$
 because it gives a contribution of order $\tilde{\alpha}^4$ to the action.

Other BPS equations are
 \beqn 
 &&\frac{f'}{r}=\frac{e_0^2}{2} 
\left( 2 \kappa^2  z^2 +\frac{\tilde{\alpha}^2}{4} \kappa^2 (z^2-1)^2 -2\right),
 \nonumber\\[3mm]
 && \frac{g'}{r}- (1+f_3) (1-g) \frac{z^2-1}{z^2+1} 
 \frac{1}{r^2}=\frac{e_3^2}{2} \kappa^2 z(z^2-1)\,, 
\nonumber\\[3mm]
 && \frac{f}{r}=2\frac{\kappa'}{\kappa}+2\frac{z'}{z} \,,
 \nonumber\\[3mm] 
 &&\frac{1-g}{r}= \frac{1}{2} \frac{z^2+1}{2z}
 \left(\frac{1}{r}- \frac{z'}{z}\right)  .
\label{laida}
\eeqn

In what follows we put $\tilde{\alpha}=0$ in 
in the first equation in (\ref{laida}).
The following change of variables is used:
\beqn 
\kappa &=&\frac{w}{\phi^2}, \,\,\,\,\, \,\,\,  z=\phi^2\,,
\nonumber\\[3mm]
\phi &=& \sqrt{z_1}=\sqrt{z_2}, \,\,\, w=\kappa z \,.
\eeqn
In these variables our problem reduces to
\beq 
f=2  \frac{w'}{w} r, \,\,\,\,\, f'=e_0^2 r (w^2-1)\, .  
\eeq
These  are  equations for the (1,1) Abelian vortex,
see Eq.~(\ref{baba}) for $p=k=1$. In adddition, we
have  new profile functions which  satisfy
the  equations
\beqn 
 && \frac{1-g}{r}= \frac{1}{2} \frac{\phi^4+1}{\phi^2}
 \left(\frac{1}{r}-2 \frac{\phi'}{\phi}\right),
\nonumber\\[3mm]
 && \frac{g'}{r}- (1-g) \frac{\phi^4-1}{\phi^4+1} 
 \frac{1}{r^2}=\frac{e_3^2}{2} (\phi^4-1) \frac{w^2}{\phi^2}\, .
\eeqn
Let us rewrite them in a form convenient for numerical
calculations, 
\beqn
 && \phi'=\frac{\phi}{2 r} \left( 1 - (1-g) \frac{2 \phi^2}{1+\phi^4} \right)\,
\nonumber\\[3mm] 
 && g'= (\phi^4-1) \left\{ \frac{e_3^2}{2} \frac{w^2 r}{\phi^2} + \frac{1-g}{r (1+\phi^4)} \right\}.
\eeqn

The squark field  can be written as
\beqn
 Q = \left(\begin{array}{cc} 
w e^{i \varphi}& -\frac{\tilde{\alpha}}{2} 
 (e^{2 i \varphi} \phi^2 w - e^{i \varphi} w) \\[3mm]
-\frac{\tilde{\alpha}}{2}  ( e^{i \varphi} w 
-\frac{w}{\phi^2}) & w e^{i \varphi} \\
\end{array}\right). 
\label{pluto}
\eeqn
The expression for the gauge field in non-singular gauge
is completely straightforward, see Eq. (\ref{f1}), (\ref{f2}).
The profile function $h$ is given by
\beq 
h = \frac{\phi^4 -1}{\phi^4 + 1} (1-g)\,.
\eeq
Now, let us compute the value of the gauge invariant
operator $ \tilde{\mathcal{F}}^a $ at   first order in $\tilde{\alpha}$,
\beqn 
\tilde{\mathcal{F}}^3 &=&  0 \,,
\nonumber\\[3mm] 
 \tilde{\mathcal{F}}^1 &=& 2 \tilde{\alpha} w^2 (\cos\varphi) 
 \left(\frac{g'}{r}-\frac{h}{r^2}\right) \,,
\nonumber\\[3mm] 
 \tilde{\mathcal{F}}^2 &=& -2  \tilde{\alpha} w^2  (\sin\varphi)
 \left(\frac{g'}{r}-\frac{h}{r^2}\right) \,.
 \label{topolino} 
 \eeqn

In particular, Eqs. (\ref{pluto}) and (\ref{topolino})
give us the Abelian (1,1) vortex at $\tilde{\alpha}=0$ (note that
the 2-string has also the U(1) gauge field $F^0_{12}=-f'/r$).

Let us consider the action of a global color+flavor
rotation, given by an SU(2) matrix $U$
\[   \mathcal{F}^a \tau^a \rightarrow  U\,\mathcal{F}^a \tau^a\, U^{\dagger},
 \,\,\,Q \rightarrow  U\,Q\, U^{\dagger}.
\]
The action is trivial only at $\tilde{\alpha}=0$;
otherwise the situation is similar to a rotation of a rigid body in
the ordinary 3-dimensional space. All  $SU(2)$ global generators act
non-trivially  on the solution (\ref{topolino}),(\ref{pluto}).
At fixed $\tilde{\alpha}$ our solution is  parameterized   by
some kind of the Euler angles in color space.
The ``shape" in color space is similar to a coin
with vanishing thickness (at the leading order in
 $\tilde{\alpha}$) and with diameter
of the order of $\tilde{\alpha}$.

\section*{Appendix B: Zero modes for the \boldmath{$(2,0)$} vortex}
\renewcommand{\theequation}{B.\arabic{equation}}
\setcounter{equation}{0}

Now we consider a small perturbation around the $(2,0)$ vortex
at $\alpha=0$. Again, acting in the same way as in  the case of 
the (1,1) string,
we calculate our profile functions with accuracy which ensures 
cancellation of the
 first and second order 
corrections with respect to 
 $\alpha$ in the action, see  Eq. (\ref{azione}).
 As before, at this order it is consistent to treat $\kappa,z_1,z_2$
as constants in $\alpha$. We also  calculate $g,h$ at order 
$\mathcal{O}(1)$ in $\alpha$.
On the other hand, we need to consider  $\mathcal{O}(\alpha^2)$ 
corrections to the functions $f,f_3$.

The BPS equations are
\beqn
&& \frac{f'}{r}=\frac{e_0^2}{2} 
(\kappa^2 +\kappa^2 z_1^2 z_2^2 -\frac{\alpha^2}{4}\kappa^2(1-z_1^2)(1-z_2^2) -2)\,,
\label{locomotiva} \\[3mm] 
&& \frac{f'_3}{r}=\frac{e_3^2}{2} 
(z_1^2 z_2^2 -1) \kappa^2 -\frac{\alpha^2}{8}\kappa^2 (z_1^2+1)(z_2^2-1)\,,
\\[3mm] 
 &&\frac{g'}{r}+\frac{1-f_3}{r^2} \frac{z_1^2-1}{z_1^2+1} (1-g)
=\frac{e_3^2}{2} \kappa^2 z_1 (z_2^2-1) \,,
\label{locomotiva3}
\\[3mm] 
&& \frac{f}{r}= 2\frac{\kappa'}{\kappa}+\frac{z_1'}{z_1}+\frac{z_2'}{z_2} \,,
\\[3mm]
&&\frac{f_3}{r}=  \frac{z_1'}{z_1}
+\frac{z_2'}{z_2} \,,
\\[3mm]
&& \frac{1-g}{r}=  \frac{z_1^2+1}{2 z_1}
 \left(\frac{1}{r}- \frac{z_2'}{z_2}\right) .
\eeqn
Instead of $z_1$, $z_2$ and $\kappa$ we introduce new profile functions,
\beqn 
z_1&=& \frac{s}{t \phi^2}, \,\,\,\,\,\, z_2=\phi^2, \,\,\,\,\,\,  \kappa=t \,, 
\nonumber\\[3mm] 
 s &=&\kappa z_1 z_2, \,\,\,  \,\,\, t=\kappa ,  \,\,\,\,\,\, \phi=\sqrt{z_2}\,. 
\eeqn
With this change of variables we find 
the following equations:
\beqn
 \frac{f'}{r}
 &=&
 \frac{e_0^2}{2} 
(s^2+ t^2-2)\, ,
\,\,\,\,\,\,\,  \frac{f'_3}{r}=\frac{e_3^2}{2} 
(s^2- t^2)\,,
\nonumber\\[3mm]
 \frac{f}{r}
 &=&
  \frac{s'}{s}+\frac{t'}{t}, \,\,\,\,\,\,\, 
 \frac{f_3}{r}= \frac{s'}{s}-\frac{t'}{t}\,.
 \eeqn
These equations coincide with the first order equations (\ref{baba})
for the Abelian (2,0) string ($p=2$, $k=0$). They can be solved separately.

 Equations for the  zero mode profile functions have the form
\beqn 
&& \frac{1-g}{r}= \frac{1}{2} \frac{s^2+t^2 \phi^4}{s t \phi^2}
 \left(\frac{1}{r}-2 \frac{\phi'}{\phi}\right) ,
\nonumber\\[3mm]
&& \frac{g'}{r}+\frac{1-f_3}{r^2} \frac{s^2-t^2 \phi^4}{s^2+t^2 \phi^4} (1-g)
=\frac{e_3^2}{2} (\phi^2-\frac{1}{\phi^2}) \, s\,  t\,.
\eeqn
Let us rewrite them in a form convenient for numerical
calculations,
\beqn
&& \phi'=\frac{\phi}{2 r} \left( 1 - (1-g) \frac{2 s t \phi^2}{s^2+t^2 \phi^4} \right) ,
\nonumber\\[3mm] 
&& g'=  \frac{e_3^2}{2} \frac{\phi^4-1}{\phi^2} r s t  -  \frac{(1-g)(1-f_3)}{r }  \frac{s^2-t^2 \phi^4}{s^2+t^2 \phi^4}\,.
\eeqn
Numerical solutions can be found, see Sects.~\ref{numerical}
and \ref{physical} (for numerical studies we take
 $e_0^2=1$ and $e_3^2=2$).

Furthermore, the squark field  can be written as
\beq
 Q = \left(\begin{array}{cc}
s e^{2 i \varphi}& \frac{\alpha}{2} (e^{2 i \varphi} s - e^{i \varphi} \frac{s}{\phi^2}) 
\\[3mm]
\frac{\alpha}{2} ( e^{i \varphi} t \phi^2 -t) & t \\
\end{array}\right).
\eeq
The profile function $h$ is given by
\beq 
h = \frac{s^2-t^2 \phi^4 }{s^2+t^2 \phi^4 } (1-g)\,.
\eeq 
Calculating  the value of the  gauge invariant
operator $\tilde{\mathcal{F}}^a$
at  first order in $\alpha$ we obtain
\beqn
 \tilde{\mathcal{F}}^3 
 &=&
  \frac{f_3'}{r} (s^2 + t^2) \,,
 \nonumber\\[3mm]
\tilde{\mathcal{F}}^1  
&=&
  \alpha (s^2+t^2) \frac{f_3'}{r}  +
\alpha (\cos \varphi)
 \left\{ 2 \left(\frac{g'}{r}+\frac{h(1-f_3)}{r^2}\right) \, s\,  t
 -\frac{f_3'}{r}\left(\frac{s^2}{\phi^2}+t^2 \phi^2\right) \right\} ,
 \nonumber\\[3mm]
\tilde{\mathcal{F}}^2 
 &=&
-
\alpha (\sin \varphi)
 \left\{ 2 \left(\frac{g'}{r}+\frac{h(1-f_3)}{r^2}\right) \,s\, t
 -\frac{f_3'}{r}\left(\frac{s^2}{\phi^2}+t^2 \phi^2\right) \right\}. 
 \eeqn
 It is possible to  globally rotate the solution,
so that $\mathcal{F}^{(1,2)}$ have no constant parts, and
their average with respect to  $\varphi$ vansihes (a``minimal'' solution). 
The matrix which realizes this transformation has the form
\beq 
\tilde{U}=  \exp(i \alpha \frac{\tau_2}{2}),
\eeq 
 acting on the fields as
 \beq
   \mathcal{F}^a \tau^a \rightarrow 
 \tilde{U}^{\dagger}\,\mathcal{F}^a \tau^a \,\tilde{U}, \,\,\,
 \,\,\,Q \rightarrow  \tilde{U}^{\dagger}\,Q \,\tilde{U}.
\eeq
The result of the rotation is 
  \beqn
   \tilde{\mathcal{F}}^3 &=& \frac{f_3'}{r} (s^2 + t^2) \,,
\nonumber\\[3mm]
 \tilde{\mathcal{F}}^1 &=& 
2 \alpha (\cos\varphi)
 \left\{ 2 \left(\frac{g'}{r}+\frac{h(1-f_3)}{r^2}\right) s\, t
 -\frac{f_3'}{r}\left(\frac{s^2}{\phi^2}+t^2 \phi^2\right) \right\} 
 \nonumber\\[3mm]
 \tilde{\mathcal{F}}^2 &=& -
2 \alpha (\sin\varphi)
 \left\{ 2 \left(\frac{g'}{r}+\frac{h(1-f_3)}{r^2}\right) s\, t
 -\frac{f_3'}{r}\left(\frac{s^2}{\phi^2}+t^2 \phi^2\right) \right\}
\\[4mm]
 Q &=& \left(\begin{array}{cc}
s e^{2 i \varphi}& \frac{\alpha}{2} (-e^{ i \varphi} \frac{s}{\phi^2} +  t) \\[3mm]
\frac{\alpha}{2}   ( -e^{2i \varphi}s+ e^{ i \varphi} t \phi^2 ) & t \\
\end{array}\right). 
\label{topolino2}
 \eeqn

At $\alpha=0$ these equations give us a solution for the Abelian
(2,0) string, see (\ref{astring}).
At $\alpha=0$ the action of the global $SU(2)$
 is similar to the   rotation of a stick
 of zero thickness in the three-dimensional space:
 the moduli space is $S^2 \approx \mathbb{CP}^1$.
 At $\alpha \neq 0$ the situation is similar to 
a rigid body rotation in the
ordinary three-dimensional space. All $SU(2)$ global generators act
non-trivially  on the solution (\ref{topolino2})).
At fixed $\alpha$ our solution is  parameterized   by
 the Euler angles in color space.
The ``shape" in color space is similar to a cigarette
with  thickness  of order $\alpha$ and length $2$.

\end{document}